\documentclass[namedreferences]{solarphysics}

\usepackage[hyperref,optionalrh,solaromanenum]{spr-sola-addons} 

\usepackage{graphicx}
\usepackage{amssymb}                    
\usepackage{lscape}
\usepackage{sidecap}
\usepackage{color}                       
\usepackage{breakurl}                         

\chardef\us=`\_

\begin{document}
\begin{article}
\begin{opening}
\title{Dependence of North--South  Difference in the Slope of  Joy's law
 on Amplitude of Solar Cycle} 

\author{\fnm{J.}~\lnm{Javaraiah}\orcid{0000-0003-0021-1230}}

\institute{Bikasipura, Bengaluru-560 111,  India.\\
Formerly working at Indian Institute of Astrophysics, Bengaluru-560 034, India.\\
email: \url{jajj55@yahoo.co.in;  jdotjavaraiah@gmail.com; jj@iiap.res.in}\\
}

\runningauthor{J. Javaraiah}
\runningtitle{North--South Asymmetry in Joy's law}

\begin{abstract}
Study of the tilt angles of solar bipolar magnetic regions 
is important because the 
tilt angles have an important role in the solar dynamo. We analysed the data 
on tilt angles of sunspot groups measured at the 
  Mt. Wilson Observatory (MWOB) during the period 1917\,--\,1986 and 
  Kodaikanal Observatory (KOB) during the period 1906\,--\,1986. 
 We binned the daily tilt-angle data
during each of the Solar Cycles 15\,--\,21  into different 
$5^\circ$-latitude intervals and calculated the mean value of the tilt angles 
  in each latitude interval and the corresponding standard error. We fitted 
these binned data to Joy's law (increase of the tilt angle with latitude), i.e. 
the linear relationship between tilt angle and  latitude of an active region.
The linear-least-square fit calculations were done by taking into account 
the uncertainties  in both the abscissa (latitude) and
 ordinate (mean tilt angle). The calculations were carried out
  by using both the tilt-angle and  area weighted tilt-angle data in  
 the whole sphere, northern hemisphere,  and southern hemisphere 
 during the  whole period  and  during each individual solar cycle.
We find a significant
 difference (north--south asymmetry) between the slopes of Joy's law
 recovered from northern and southern hemispheres' whole period MWOB data 
of area-weighted tilt angles. 
 Only the slope  obtained from  the southern hemisphere's MWOB data of
 a solar cycle
 is found to be  reasonably well anti-correlated to the amplitude of the solar
 cycle.  In the case of area weighted tilt-angle data, a good correlation
 is  found between the  north--south
 asymmetry   in the slope of a solar cycle and the  amplitude of
the  solar cycle. 
The corresponding best-fit linear equations are  found to be statistically
 significant.
\end{abstract}

\end{opening}

\section{Introduction}
The tilt angles of bipolar  magnetic regions 
 are one of the important ingredients for solar-dynamo models  
\citep[][and references therein]{bhowmik18}. Tilt 
angles have a significant role in the generation of polar magnetic fields and 
reverse of polarity of  polar fields in a 22-year solar magnetic cycle.  
The  tilt angles and Joy's law (increase of tilt  angle with latitude) depend
 on several
 properties of solar magnetic active regions (sunspot groups), e.g.  latitude,  area, 
 formation and evolution, rotation and meridional motions, and also on
 solar-cycle phase~\citep{howard91,siva99,muneer02,zharkova08}.
\citet{dasi10} analysed 
Mt. Wilson Observatory  (1917\,--\,1986) and Kodaikanal Observatory
 (1906\,--\,1986)  sunspot-group tilt-angle 
data and found the existence of a reasonably good anti-correlation between 
the ratio of mean  tilt angle  [$\langle \gamma \rangle$] over a solar cycle 
 to  mean  absolute latitude
 [$\langle |\lambda| \rangle$] over the same solar cycle  and 
the strength/amplitude of
 the solar cycle  (note that  `$\langle . \rangle$'
implies the mean over a time interval, i.e. here over a solar cycle).
 Some  authors confirmed this result  and  some others 
criticized  it for various reasons~\citep{jiao21}.
 There  exists a considerable  north--south 
difference/asymmetry in strengths/amplitudes of many solar cycles 
\citep[e.g.][]{cob93,verma93,ng10,mci13,jj19,rcj21}. North--south asymmetry 
also exists  in the rotational  and  the meridional 
motions of solar plasma,  magnetic field, and tracers such as sunspot groups
  \citep{hw90,jg97,haber02,jj03,gigo05,ju06,xie18,lekshmi18,wan22}. 
 It is believed that the action of Coriolis force on rising magnetic  flux tubes
 responsible for the tilts of bipolar active regions and Joy's law
 \citep[see][]{dsilva93,fisher95,bhowmik18,jiao21}. 
Since there exists north--south asymmetry in the solar rotation, there 
could be a difference in the action of Coriolis forces on the rising 
magnetic-flux tubes in northern and southern hemispheres. Therefore, 
north--south asymmetry may  exist in the  slope  
of Joy's law  and it also may depend on amplitude of the solar cycle.
\citet{norton13} 
analysed the Mt. Wilson sunspot-group data (1917\,--\,1986) and
 found  the aforementioned 
result (anti-correlation between 
$\langle \gamma \rangle/\langle |\lambda| \rangle$  
and strength of cycle)   from the southern hemisphere's data 
 and not found from the northern hemisphere's data. 
  \citet{dasi10} and \citet{norton13} used  
$\langle \gamma \rangle/\langle |\lambda| \rangle$  
 to remove   
the  effect of latitudinal dependence of  mean tilt angle in  
the relation between the mean tilt angle and the strength of the solar cycle.
However, the anti-correlation between $\langle \gamma \rangle$ alone and
 strength of a solar cycle is statistically  insignificant. Therefore, one
 may have a doubt that    the existence of a good correlation between  
 $\langle |\lambda| \rangle$ and the strength/amplitude of a solar cycle    
 have  a significant influence on the aforementioned result. 
That is,   the significant anti-correlation between
 $\langle \gamma \rangle/\langle |\lambda| \rangle$  
 and the strength/amplitude of a solar cycle 
 might be largely an artifact of the good correlation 
between the denominator [$\langle |\lambda| \rangle$] of
 $\langle \gamma \rangle/\langle |\lambda| \rangle$
 and the strength/amplitude of the solar cycle.
It is difficult to differentiate   the coefficients of Joy's law
 (namely the slope of the  linear relation between tilt angle 
and latitude) of 
different solar cycles due to large uncertainties in the derived
  coefficients \citep{dasi10}.  Hence,  \cite{dasi10}
and \citet{norton13} did not calculate correlation between the slope 
of Joy's law and amplitude/strength of solar cycle. 
 In the present analysis, by using the same aforementioned  
data of  tilt angles 
of the sunspot groups measured in
 Mt. Wilson Observatory (MWOB) during the period 1917\,--\,1986 and 
Kodaikanal Observatory (KOB) during the period 1906\,--\,1986, 
 we study the dependence of the slope of Joy's law  (coefficient of Joy's law)
 on the
amplitude of the solar cycle  by determining it  from the whole sphere's  
 data and 
  northern and southern hemispheres' data. 
 We obtain the relationship between the slope 
 (and its north--south asymmetry)  
  and the amplitude of the solar cycle
 by determining the linear-least-square 
fits to the data of these parameters  taking into account the
 uncertainties in  all these parameters.     
 
In the next section we describe the data and analysis. In Section~3 
we describe the results, and in Section~4 we present the conclusions 
and briefly discuss them.

\section{Data Analysis}
Here we  use  the daily  data: heliographic latitude 
 (area weighted) [$\lambda$],  area [$A$],  tilt angle [$\gamma$], etc. of
 sunspot groups measured
 in  Mt. Wilson 
Observatory (MWOB) during the period 1917\,--\,1986, and 
Kodaikanal Observatory (KOB) during the period 1906\,--\,1986. 
  These data are available at
the website {\sf www.ngdc.noaa.gov\break/stp/solar/sunspotregionsdata.html}.
We had used these data
  in an our  earlier analysis of angular velocities of 
sunspot groups~\citep{jbu05}. 
In both the northern and southern hemispheres a positive tilt angle
implies that the leading spot is closer to the Equator than the 
following spot.  
 We use the amplitudes $175.7 \pm 11.8$, $130.2 \pm 10.2$, 
$198.6 \pm 12.6$, $218.7 \pm 10.3$, $285.0 \pm 11.3$, $156.6 \pm 8.4$,
 and $232.9 \pm 10.2$ (values of $R_{\rm M}$, the maximum 13-month
smoothed monthly mean values of Version-2 of the 
international sunspot number, SN)
of Solar Cycles 15, 16, 17, 18, 19, 20, and 21, respectively
\citep[taken from][]{pesnell18}.
 Here the data reduction and analysis are similar as in \citet{dasi10} and 
\citet{norton13}.
We excluded the data correspond to the zero values of the tilt angles, and also 
that correspond to   
the difference between the tilt angles of a sunspot group observed in two 
consecutive days  is 
$\ge 16^\circ$ day$^{-1}$. 
 We binned the daily tilt-angle data
during each of the Solar Cycles 15\,--\,21  into seven different 
$5^\circ$-latitude (absolute)  intervals $0^\circ$\,--\,$5^\circ$, 
 $5^\circ$\,--\,$10^\circ$, $10^\circ$\,--\,$15^\circ$, 
$15^\circ$\,--\,$20^\circ$, $20^\circ$\,--\,$25^\circ$, 
$25^\circ$\,--\,$30^\circ$, and $30^\circ$\,--\,$35^\circ$.
We calculated the  mean value, [$\bar \gamma$],  of the  tilt angles, and 
the corresponding standard error
  in each latitude bin. 
We fitted these  data to  Joy's law  in the form:
$$\bar \gamma = m|\lambda| + c, \eqno(1)$$ 
where $\lambda$ is the mid-value of a latitude bin
 (in the case of area-weighted tilt angle  
$\bar \gamma_{\rm aw}$ is used  instead of $\bar \gamma$). 
Here the slope $m$ is referred to as the coefficient of Joy's law.
The calculations of linear-least-square fits  were done by taking into account 
the uncertainties in both the abscissa [$|\lambda|$] and the 
 ordinate [$\bar \gamma$], namely the standard error in the case of
 $\bar \gamma$ and the value $2.5^\circ$, i.e. half of the range of a
 latitude interval, in the case of  $|\lambda|$.
 The calculations were done by using  the   
tilt-angle data and also the area-weighted  tilt-angle data of whole sphere,
 and separately for northern and southern hemispheres' tilt-angle data.
 The  area-weighted mean tilt angle,   
 $\sum (\gamma_i \times A_i)/\sum A_i = X/Y$, where $X$ is the mean of 
($\gamma_i \times A_i$) and $Y$ is the mean of $A_i$,
 $i=1$,\dots,$n$,  and 
$n$ is the number of data points in a given latitude bin.
  If $\Delta(X)$ and $\Delta(Y)$ are the uncertainties 
in $X$ and $Y$, respectively, then the uncertainty
in the ratio $X/Y$ will be approximately equal to 
($Y \times \Delta(X)-X \times \Delta(Y)$)/$Y^2$ \citep[see also][]{jg97}.  
Note that here  $\Delta(X)$ and
$\Delta(Y)$ represent the standard errors of $X$ and $Y$, respectively. 
 The coefficients of Joy's law over the whole sphere and  
in northern and southern hemispheres were determined from the combined data
 of all cycles (15\,--\,21), and from the data of each individual cycle.
The MWOB data of Solar Cycle~15 are incomplete (note that the beginning of this 
solar cycle was 1913). In the case of this solar cycle, 
 the number of data points is found to be 
 very few in the $30^\circ$\,--\,$35^\circ$ latitude bins of
 the northern hemisphere
 (zero in both MWOB and KOB data) and the southern hemisphere (6 and 2 
in MWOB and KOB data, respectively).
 Hence,  the data in this latitude bin  
  are not considered in the linear-least-square  fits
 of this solar cycle's data 
in all three cases:
 whole sphere, northern hemisphere, and southern hemisphere.  
We determined  correlation between the slope $m$ and amplitude $R_{\rm M}$ 
during Solar Cycles 15\,--\,21. We also determined the corresponding 
linear regression. 
 The separate  northern and southern hemispheres' sunspot-number data 
are not available for Solar Cycles 
15\,--\,21. Hence, we compared the northern and southern hemispheres' 
values of the slope of 
 Joy's law and the corresponding north--south asymmetry, with the 
amplitude $R_{\rm M}$ 
(maximum total sunspot number) of the solar cycle. 
 All the linear regression analyses presented in this article 
 are done by using the Interactive Digital Library (IDL) software
 {\textsf{FITEXY.PRO}}, which  is downloaded from
 the website \textsf{idlastro.gsfc.nasa.gov/ftp/pro/math/}.
 This software is very useful because 
the uncertainties of both the abscissa and ordinate will be taken care in   
the calculations of linear-least-square fit.

\begin{table}[h]
\centering
{\tiny
\caption[]{Values of mean tilt angle [$\bar \gamma$, in deg.] and of 
mean area-weighted tilt angle [$\bar \gamma_{\rm aw}$, in deg.]  
of sunspot groups in absolute latitude intervals 
$0^\circ$\,--\,$5^\circ$, $5^\circ$\,--\,$10^\circ$,
 $10^\circ$\,--\,$15^\circ$, $15^\circ$\,--\,$20^\circ$,
 $20^\circ$\,--\,$25^\circ$, $25^\circ$\,--\,$30^\circ$, and 
$30^\circ$\,--\,$35^\circ$ 
($|\lambda|$ represents the mid-value of an absolute latitude interval) 
determined  from the whole-sphere data,  and separately from
 the northern and southern hemispheres'  data, during the entire
 period. $n$ represents the number of data points in each
 latitude interval. The  uncertainties in  $\bar \gamma$ and 
 $\bar \gamma_{\rm aw}$ are the corresponding  standard errors.}    
 \begin{tabular}{lccccccccccc}
\hline
&\multicolumn{3}{c}{Whole sphere} &\multicolumn{3}{c}{North hemisphere}&\multicolumn{3}{c}{South hemisphere}\\
$|\lambda|$ &$\bar \gamma$ &$\bar \gamma_{\rm {aw}}$&$n$&$\bar \gamma$ &$\bar \gamma_{\rm {aw}}$&$n$&$\bar \gamma$ &$\bar \gamma_{\rm {aw}}$&$n$\\
\hline
&&&\multicolumn{4}{c}{Derived from Mt. Wilson data}\\
 2.5& $ 1.44\pm 0.76$&$ 3.32\pm 0.82$& 1670&$ 0.79\pm 1.08$&$ 4.34\pm 1.34$&  820&$ 2.06\pm 1.07$&$ 2.41\pm 0.98$&  850\\
 7.5& $ 2.43\pm 0.39$&$ 3.28\pm 0.43$& 5914&$ 3.18\pm 0.56$&$ 3.78\pm 0.60$& 2845&$ 1.73\pm 0.54$&$ 2.80\pm 0.62$& 3069\\
12.5& $ 4.39\pm 0.34$&$ 5.08\pm 0.35$& 7719&$ 4.40\pm 0.46$&$ 5.21\pm 0.48$& 4108&$ 4.37\pm 0.50$&$ 4.93\pm 0.51$& 3611\\
17.5& $ 4.84\pm 0.38$&$ 5.01\pm 0.42$& 6461&$ 5.55\pm 0.53$&$ 5.75\pm 0.61$& 3310&$ 4.10\pm 0.54$&$ 4.18\pm 0.58$& 3151\\
22.5& $ 5.45\pm 0.48$&$ 6.31\pm 0.51$& 3881&$ 6.11\pm 0.64$&$ 7.69\pm 0.64$& 2190&$ 4.60\pm 0.73$&$ 4.55\pm 0.81$& 1691\\
27.5& $ 6.56\pm 0.76$&$ 7.65\pm 0.80$& 1764&$ 7.61\pm 1.02$&$ 9.58\pm 1.04$& 1051&$ 5.02\pm 1.14$&$ 4.36\pm 1.23$&  713\\
32.5& $ 6.72\pm 1.41$&$ 7.46\pm 1.57$&  544&$ 7.03\pm 1.81$&$ 8.68\pm 1.98$&  318&$ 6.28\pm 2.25$&$ 5.38\pm 2.57$&  226\\
\\
&&&\multicolumn{4}{c}{Derived from Kodaikanal data}\\
 2.5& $ 2.19\pm 0.74$&$ 3.71\pm 0.91$& 1583&$ 1.95\pm 1.01$&$ 4.84\pm 1.36$&  842&$ 2.47\pm 1.09$&$ 2.23\pm 1.14$&  741\\
 7.5& $ 2.94\pm 0.40$&$ 3.81\pm 0.44$& 5732&$ 3.65\pm 0.57$&$ 5.41\pm 0.62$& 2936&$ 2.20\pm 0.56$&$ 1.90\pm 0.60$& 2796\\
12.5& $ 4.56\pm 0.33$&$ 6.02\pm 0.36$& 7857&$ 5.38\pm 0.45$&$ 6.25\pm 0.49$& 4111&$ 3.67\pm 0.48$&$ 5.72\pm 0.53$& 3746\\
17.5& $ 4.74\pm 0.39$&$ 4.88\pm 0.41$& 6480&$ 5.31\pm 0.53$&$ 5.67\pm 0.57$& 3226&$ 4.18\pm 0.57$&$ 4.01\pm 0.60$& 3254\\
22.5& $ 5.31\pm 0.48$&$ 5.40\pm 0.55$& 4180&$ 6.38\pm 0.62$&$ 6.29\pm 0.69$& 2418&$ 3.85\pm 0.76$&$ 4.04\pm 0.92$& 1762\\
27.5& $ 7.38\pm 0.73$&$ 9.37\pm 0.87$& 1866&$ 7.12\pm 0.99$&$10.20\pm 1.30$& 1010&$ 7.69\pm 1.09$&$ 8.36\pm 1.09$&  856\\
32.5& $ 4.56\pm 1.38$&$ 6.91\pm 1.82$&  618&$ 6.68\pm 1.70$&$ 8.71\pm 2.46$&  346&$ 1.85\pm 2.28$&$ 4.28\pm 2.65$&  272\\
\hline
\end{tabular}
}
\label{table1}
\end{table}

\section{Results}
\subsection{Recovering Joy's law From the Combined Data of All Solar Cycles}
In Table~1 we have given the values of the mean 
 tilt angle [$\bar \gamma$] and of the 
mean area-weighted tilt angle [$\bar \gamma_{\rm aw}$]
of sunspot groups in each absolute latitude intervals
$0^\circ$\,--\,$5^\circ$, $5^\circ$\,--\,$10^\circ$,
 $10^\circ$\,--\,$15^\circ$, $15^\circ$\,--\,$20^\circ$,
 $20^\circ$\,--\,$25^\circ$, $25^\circ$\,--\,$30^\circ$, and
$30^\circ$\,--\,$35^\circ$
($|\lambda|$ represents the mid-value of an absolute latitude interval)
separately determined from the  data of sunspot groups in 
northern and southern hemispheres and from the combined data 
(whole sphere  data) during all solar cycles (15\,--\,21),  
i.e. data of sunspot groups  measured in MWOB during the 
 period 1917\,--\,1986 and in KOB  during the   period 1906\,--\,1986. 
 The  uncertainties in  $\bar \gamma$ and $\bar \gamma_{\rm aw}$ are 
 the corresponding standard errors and  $n$ represents
  the number of data points in a latitude interval.
As we can see in this table the values of both
  $\bar \gamma$ and $\bar \gamma_{\rm aw}$
are reasonably accurate (the ratio of mean value to standard error 
is considerably large) except, particularly in the case of KOB data, 
 at $|\lambda| = 32.5^\circ$ 
(i.e. in $30^\circ$\,--\,$35^\circ$  latitude interval) of southern 
hemisphere.
Figure~1 shows the Joy's law derived from the data given in Table~1 for  
the whole sphere and Figures~2 and 3 are the same as 
Figure~1, but the Joy's laws are derived from the 
 northern and southern hemispheres' data, respectively. In all these figures, 
 the values of the correlation coefficient [$r$], rms 
(root-mean-square deviation), $\chi^2$, and the corresponding  
probability [$P$] are given, and in Table~2 the details on the 
best-fit linear equations are given. As we can see in these figures and 
in Table~2, all the  linear relations (Joy's laws) derived from particularly 
MWOB data are good, i.e. the values of $r$ are sufficiently high, the values of 
 $\chi^2$ are significantly low, and the  ratios of the slopes to the 
corresponding standard deviations are substantially high. Overall, the relations
 derived from KOB data are relatively less reliable, especially, 
the relations correspond to both the tilt-angle  and the area-weighted 
tilt-angle data of the southern hemisphere are statistically 
insignificant (see Figures~3c and 3d). 

As we can see in Table~2, there exist 
some differences in the values of the slopes of the
 linear equations obtained from MWOB and 
KOB data. These differences are statistically significant in the case of
 area-weighted tilt-angle data of both the northern and southern hemispheres.
 The values of  the slopes determined from the MWOB and KOB  whole-sphere's 
tilt-angle data  well
 match within their uncertainties limits with the value 0.2 that was obtained by \citet{norton05} by using tilt angles of 650 active regions observed in 
{\it Michelson Doppler Imager} data during the period 1996\,--\,2004.
Both MWOB and KOB  whole sphere's values  of the slopes
(determined from both the tilt-angle and area-weighted tilt-angle data)   
 are  slightly lower than
the corresponding values $0.26 \pm 0.05$ and $0.28 \pm 0.06$  
 that were obtained from the area-weighted tilt-angle data by 
\citet{dasi10} by forcing the linear fit through the origin.  
The values of northern and southern hemispheres' slopes determined from
 MWOB area-weighted tilt-angle data  reasonably match  the  
corresponding values 0.26 and 0.13 determined
from the same data by \cite{norton13}.   
In the case of MWOB area-weighted tilt-angle  
 data, the slope  of northern hemisphere is  
significantly (more than 95\,\% confidence level)  larger ($\approx$130\,\%)
 than that of the southern hemisphere. In the case of KOB 
data, the slope corresponding to the area-weighted tilt-angle data of the
 southern
 hemisphere  looks to be larger than that of the northern 
hemisphere, but as already mentioned above, the relations obtained from 
the KOB southern hemisphere's  data are unreliable.   Hence, in this case 
 it is not possible to draw any conclusion on  the north--south difference 
in the slope.

\begin{figure}
\centering
\includegraphics[width=6.0cm]{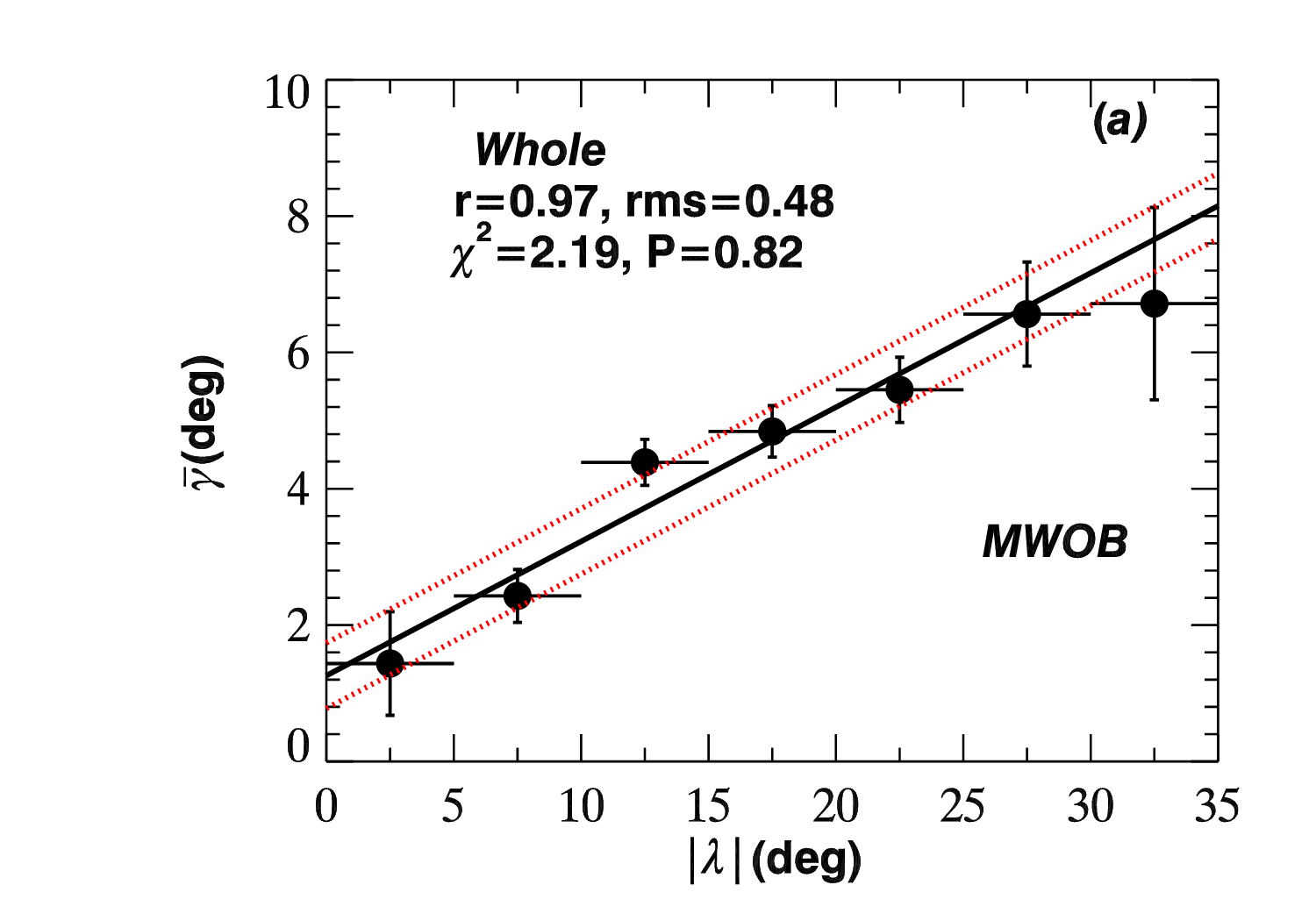}
\includegraphics[width=6.0cm]{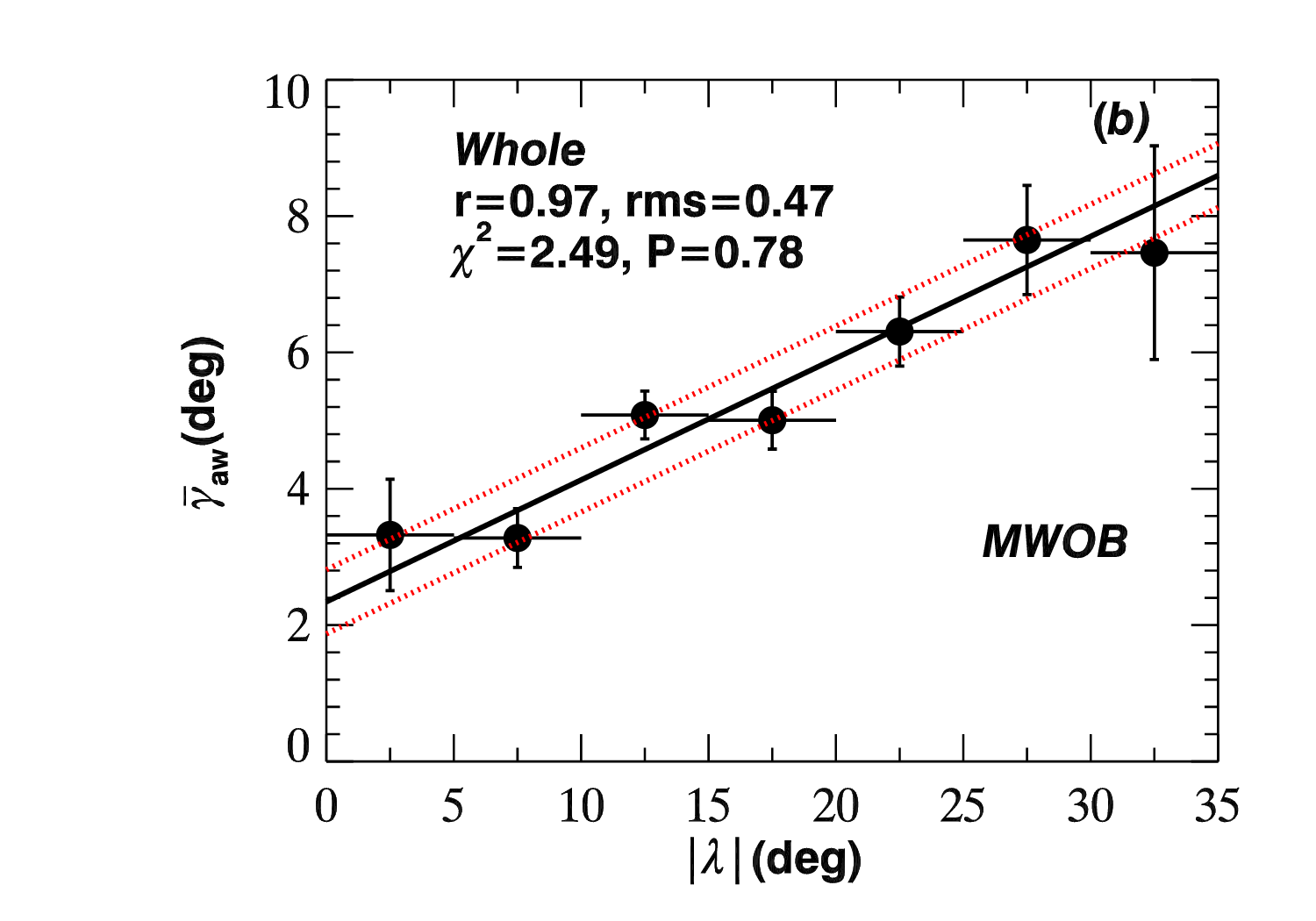}
\includegraphics[width=6.0cm]{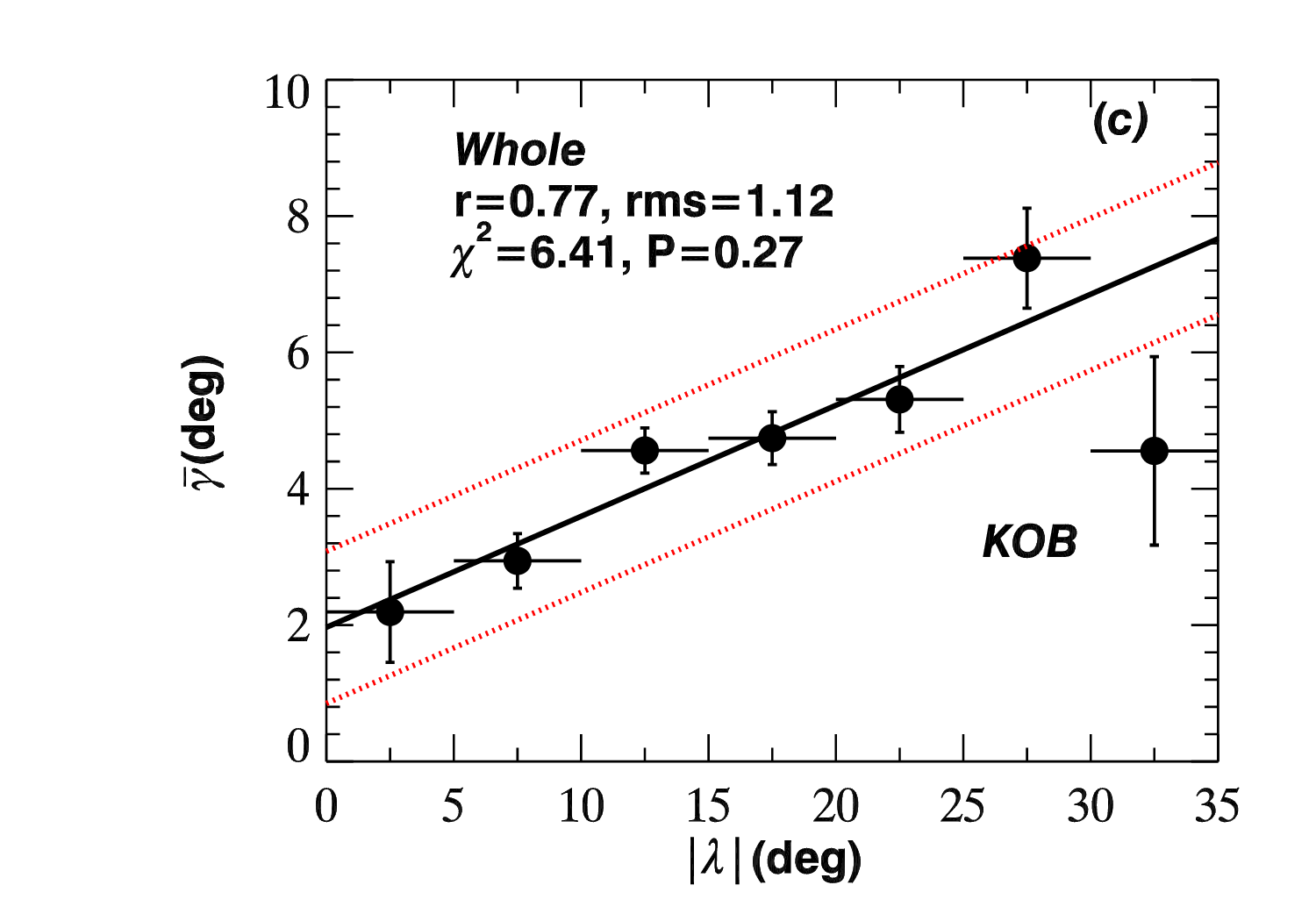}
\includegraphics[width=6.0cm]{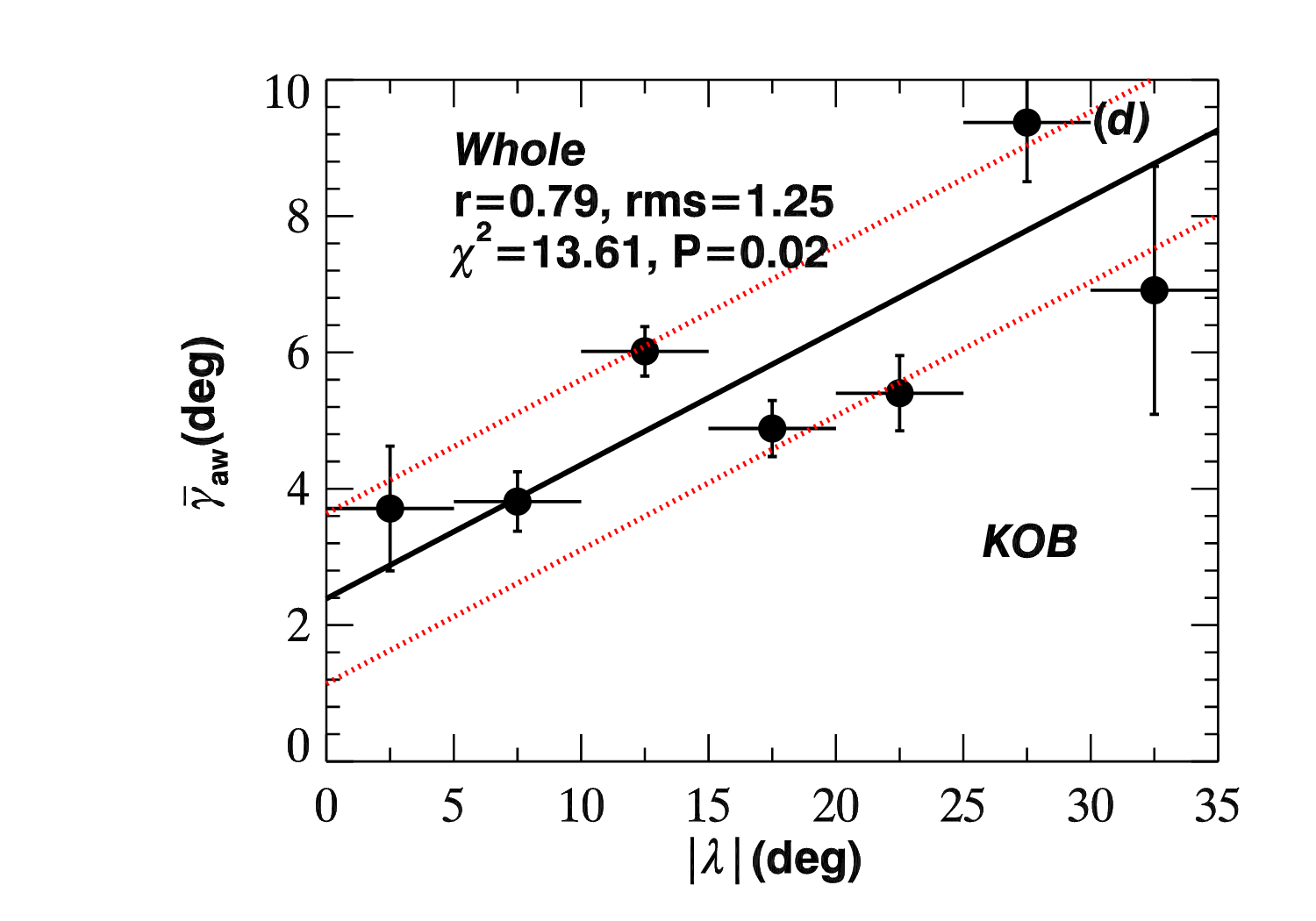}
\caption{Mean tilt angle [$\bar \gamma$] and mean 
 area-weighted tilt angle [$\bar \gamma_{\rm aw}$] versus mid-value
 of the corresponding  $5^\circ$ latitude bin, determined from 
 Mt. Wilson Observtory (MWOB)
 and Kodaikanal Observatory (KOB) whole-sphere sunspot-group data.  
The {\it continuous line} represents the corresponding  
linear-least-square best fit and the {\it dotted line} ({\it red}) represents
 one-rms level. The values of the
 correlation coefficient [$r$], rms, and $\chi^2$ and the
corresponding probability [$P$] are also given.}
\label{f1}
\end{figure}

\begin{figure}
\centering
\includegraphics[width=6.0cm]{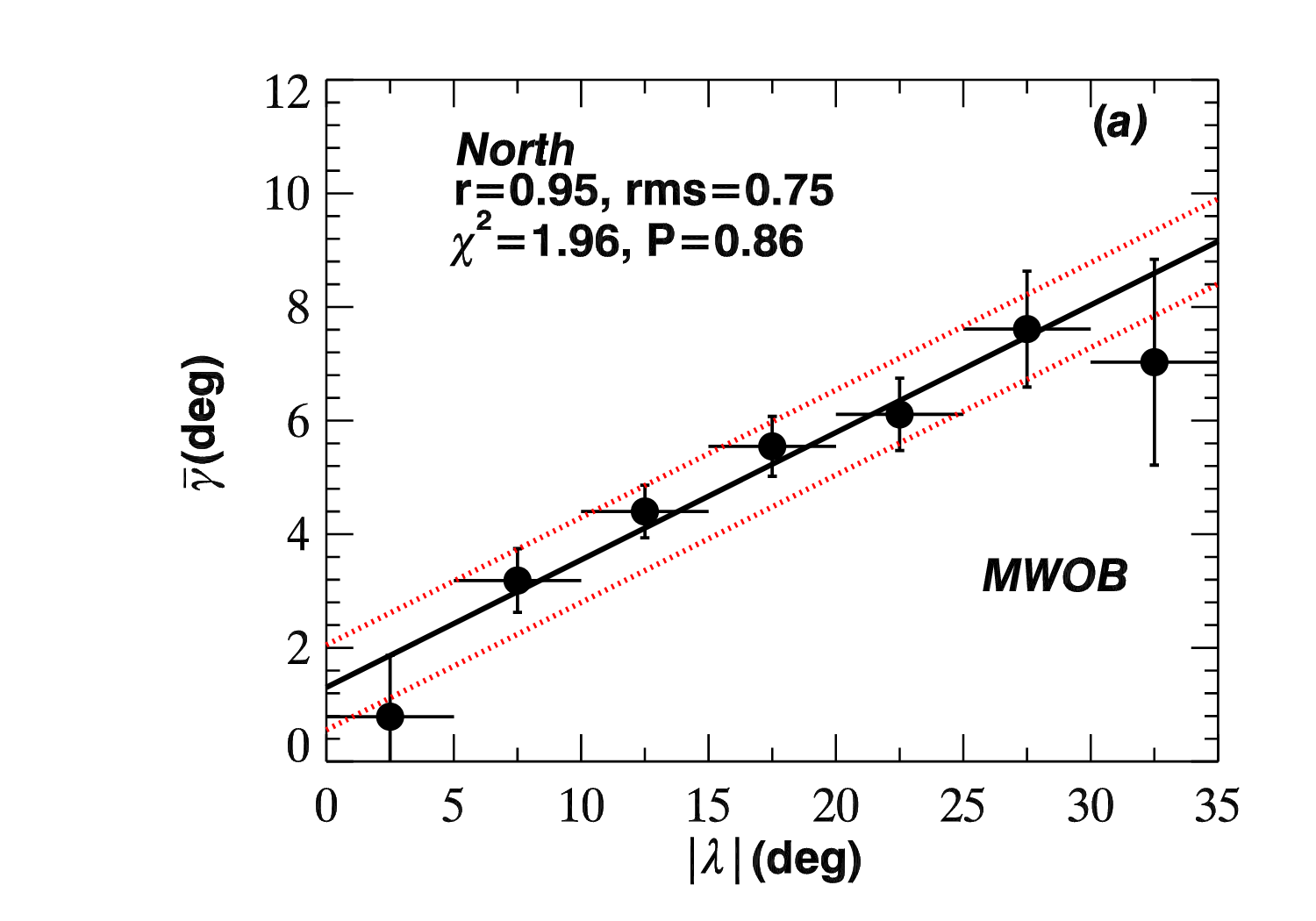}
\includegraphics[width=6.0cm]{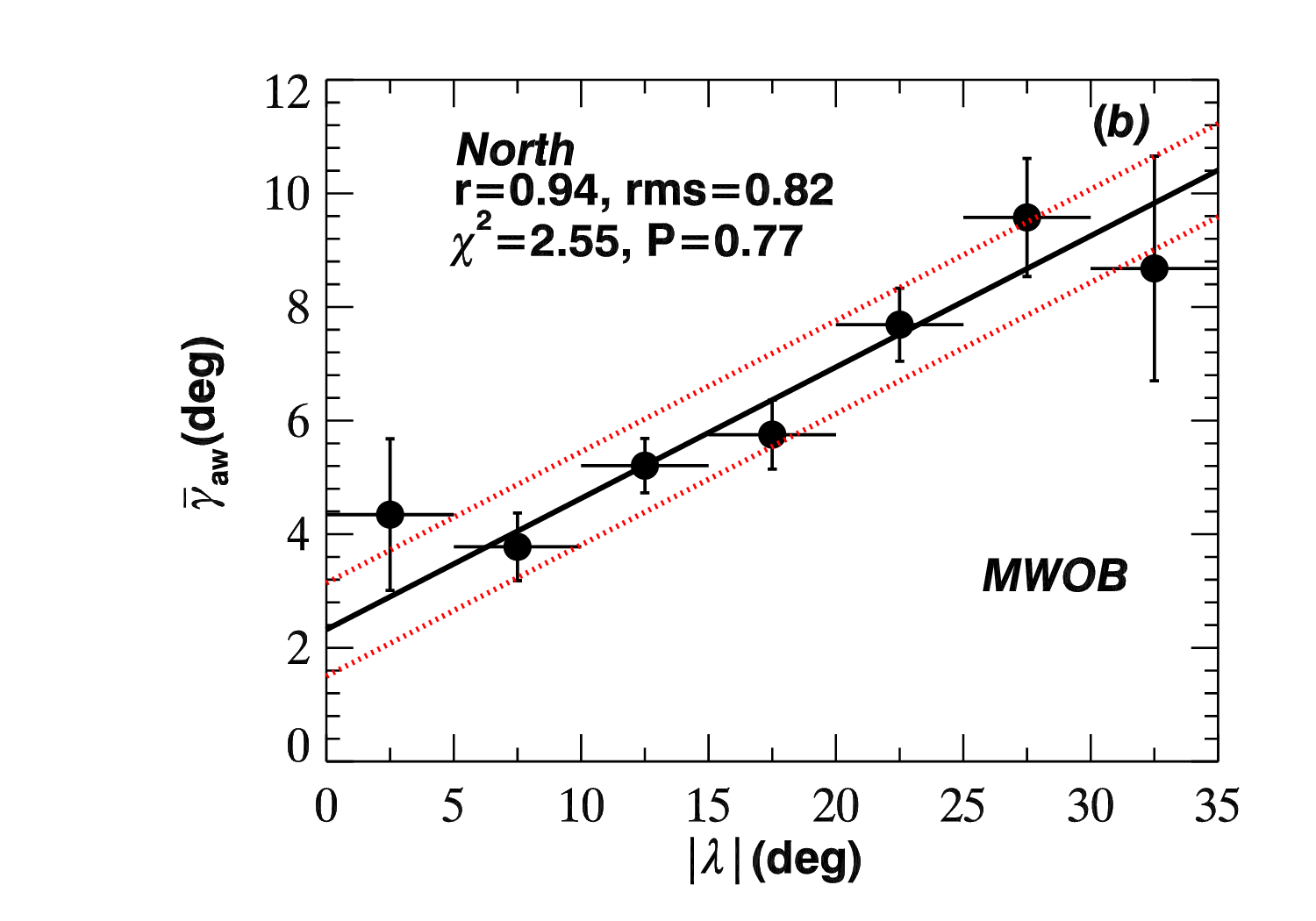}
\includegraphics[width=6.0cm]{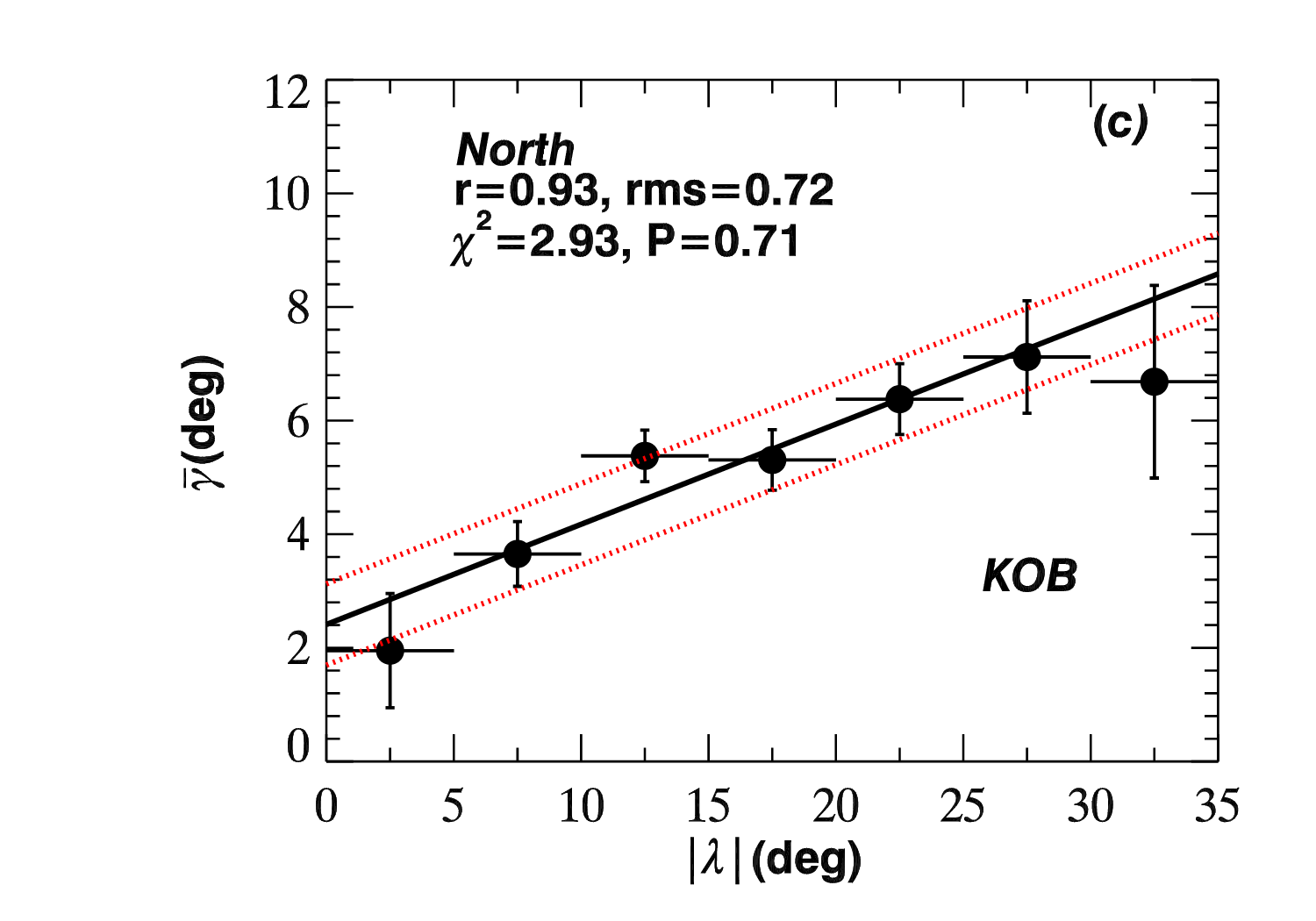}
\includegraphics[width=6.0cm]{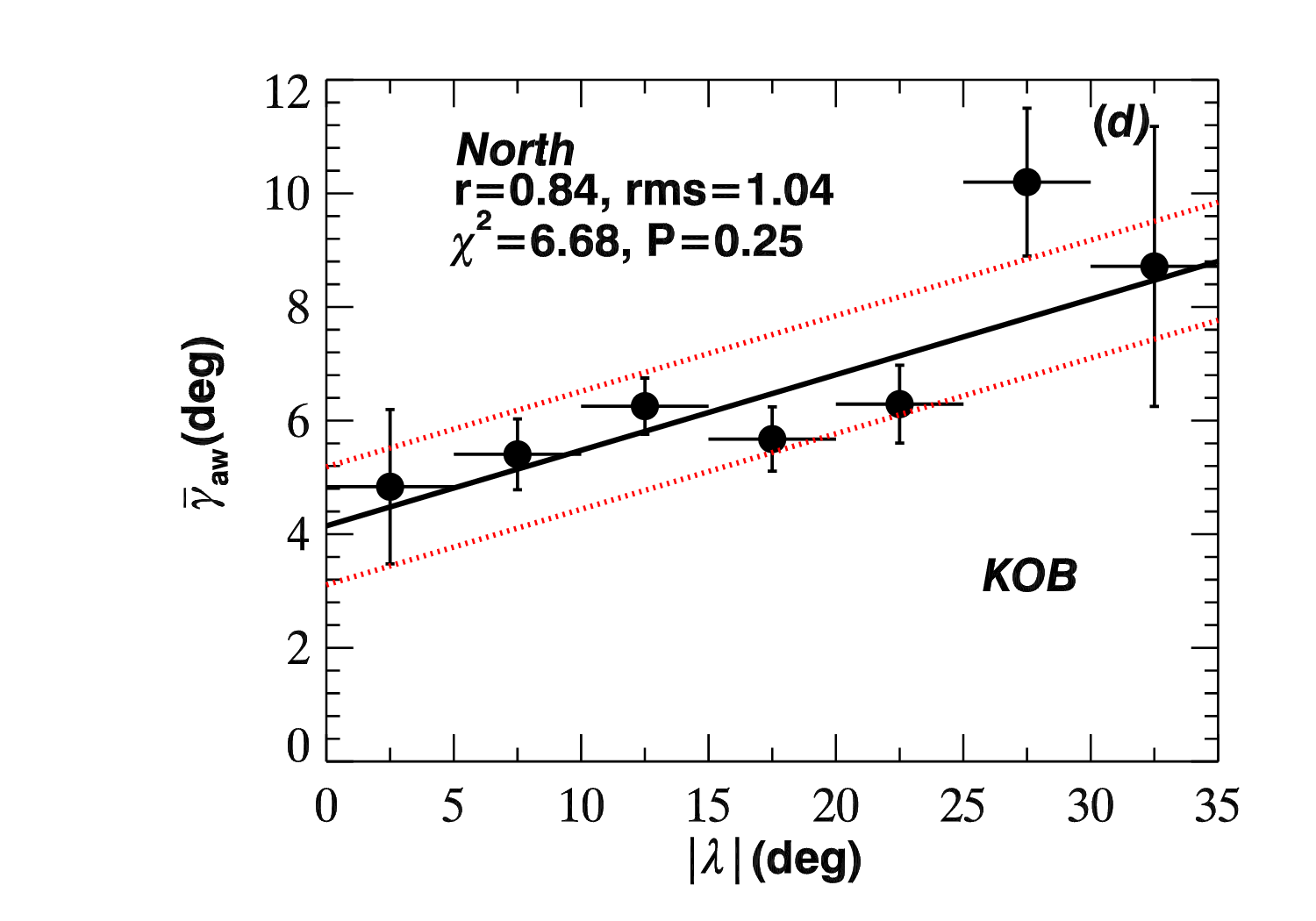}
\caption{Mean tilt angle [$\bar \gamma$] and mean 
 area-weighted tilt angle [$\bar \gamma_{\rm aw}$] versus mid-value
 of the corresponding  $5^\circ$ latitude bin, determined from 
 Mt. Wilson Observtory (MWOB)
 and Kodaikanal Observatory (KOB) northern hemisphere sunspot-group data.  
The {\it continuous line} represents the corresponding  linear-least-square
 best fit and the {\it dotted line} ({\it red}) represents one-rms level.
 The values of the
 correlation coefficient [$r$], rms, and $\chi^2$ and the
corresponding probability [$P$] are also given.}
\label{f2}
\end{figure}

\begin{figure}
\centering
\includegraphics[width=6.0cm]{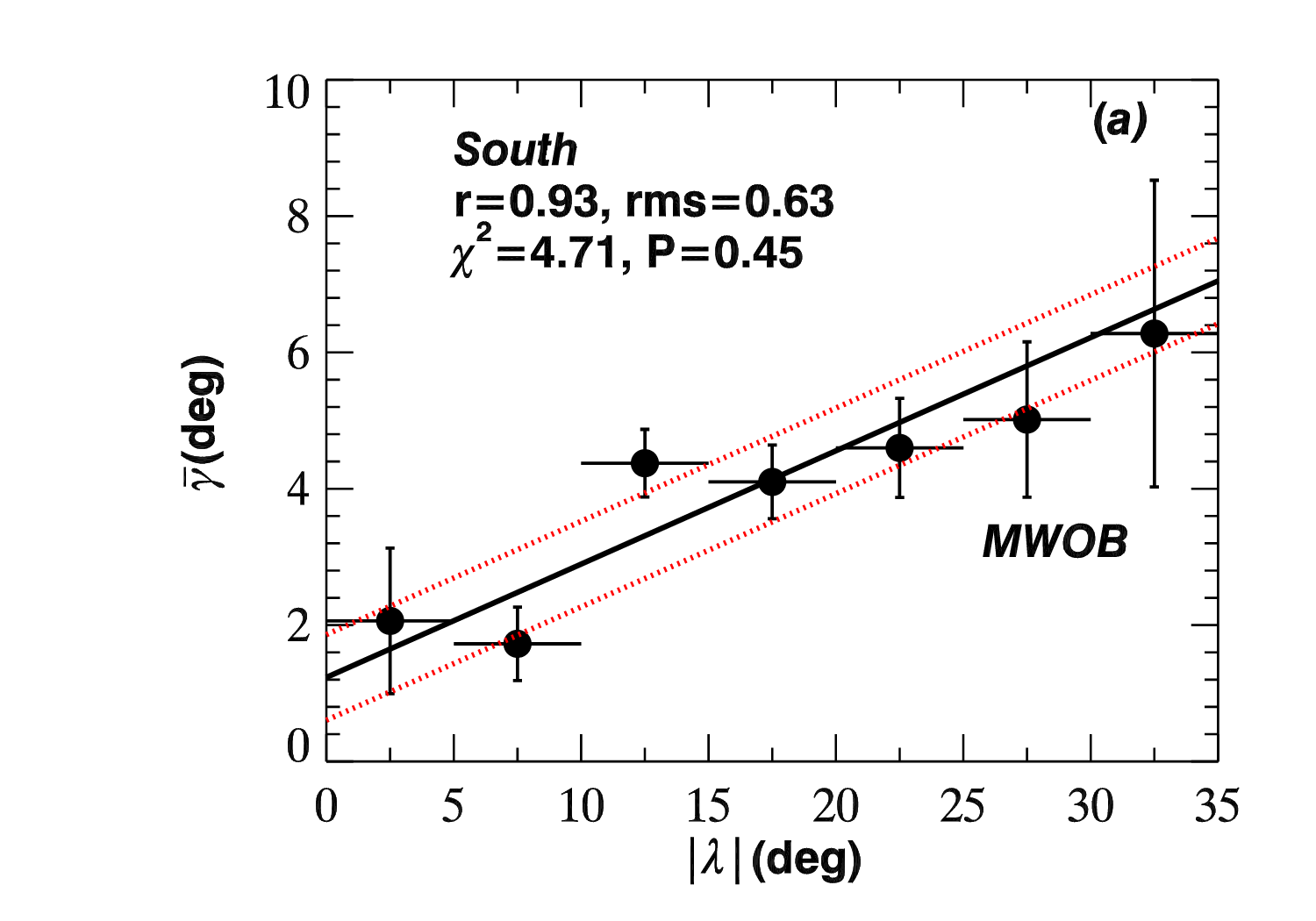}
\includegraphics[width=6.0cm]{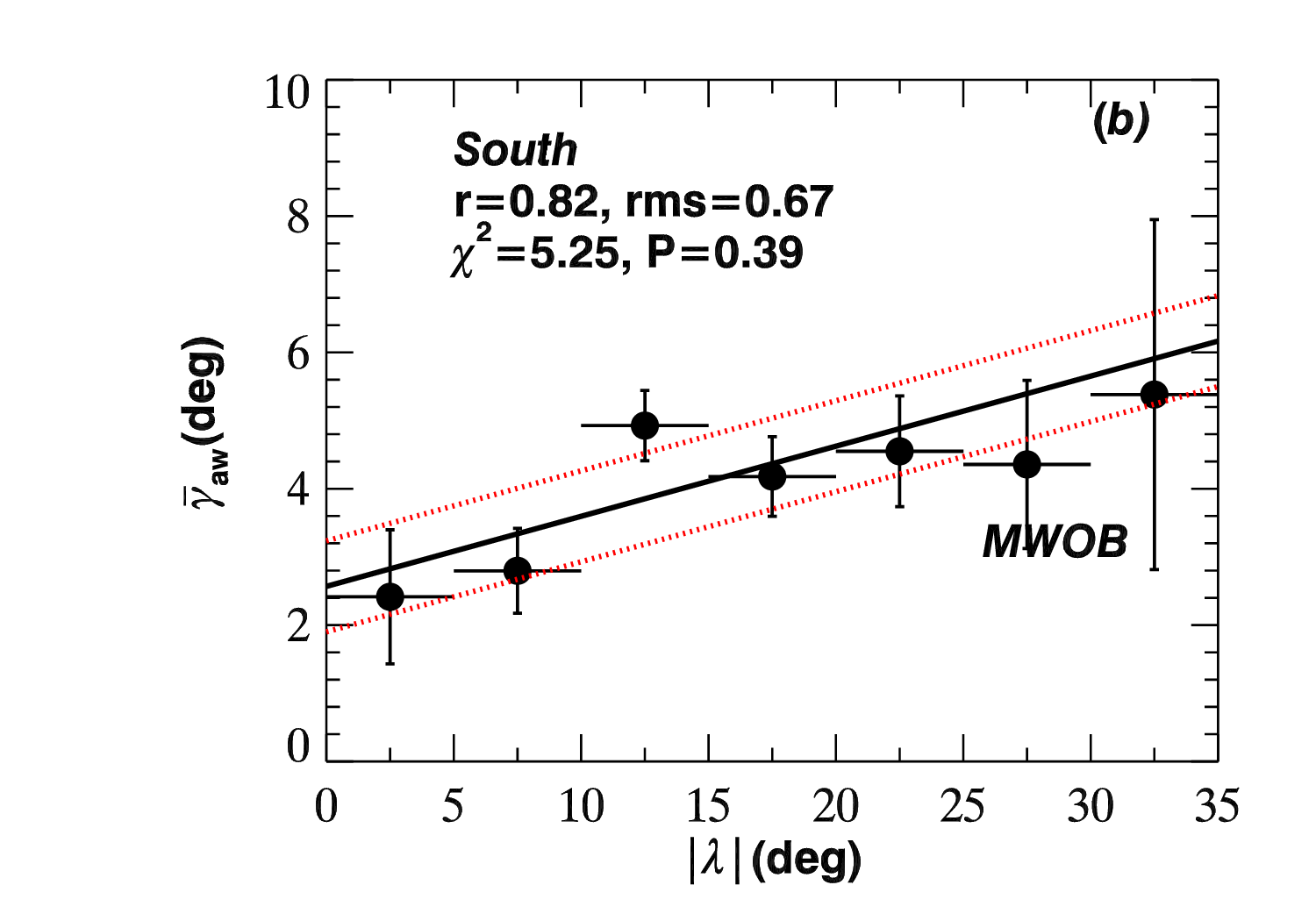}
\includegraphics[width=6.0cm]{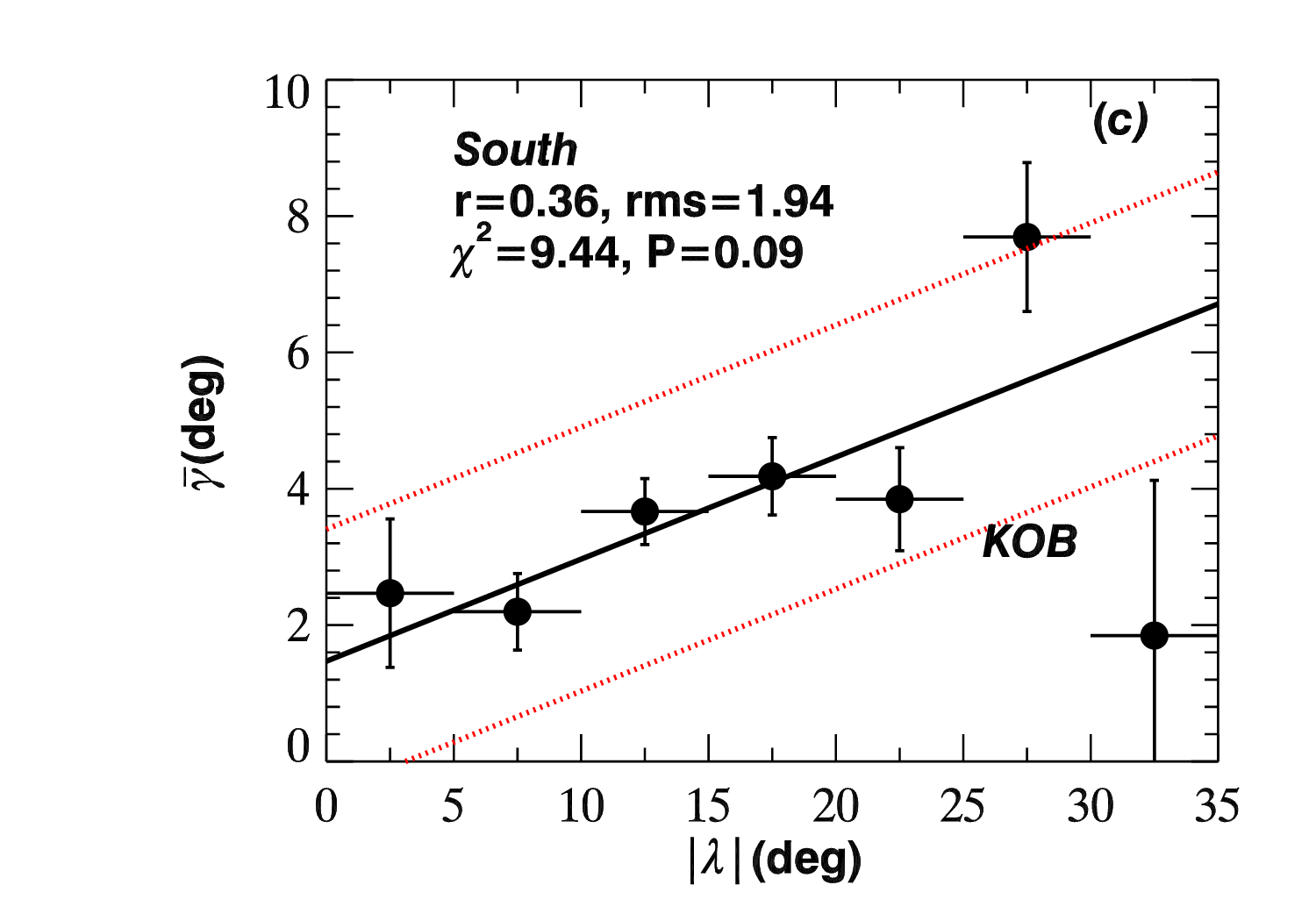}
\includegraphics[width=6.0cm]{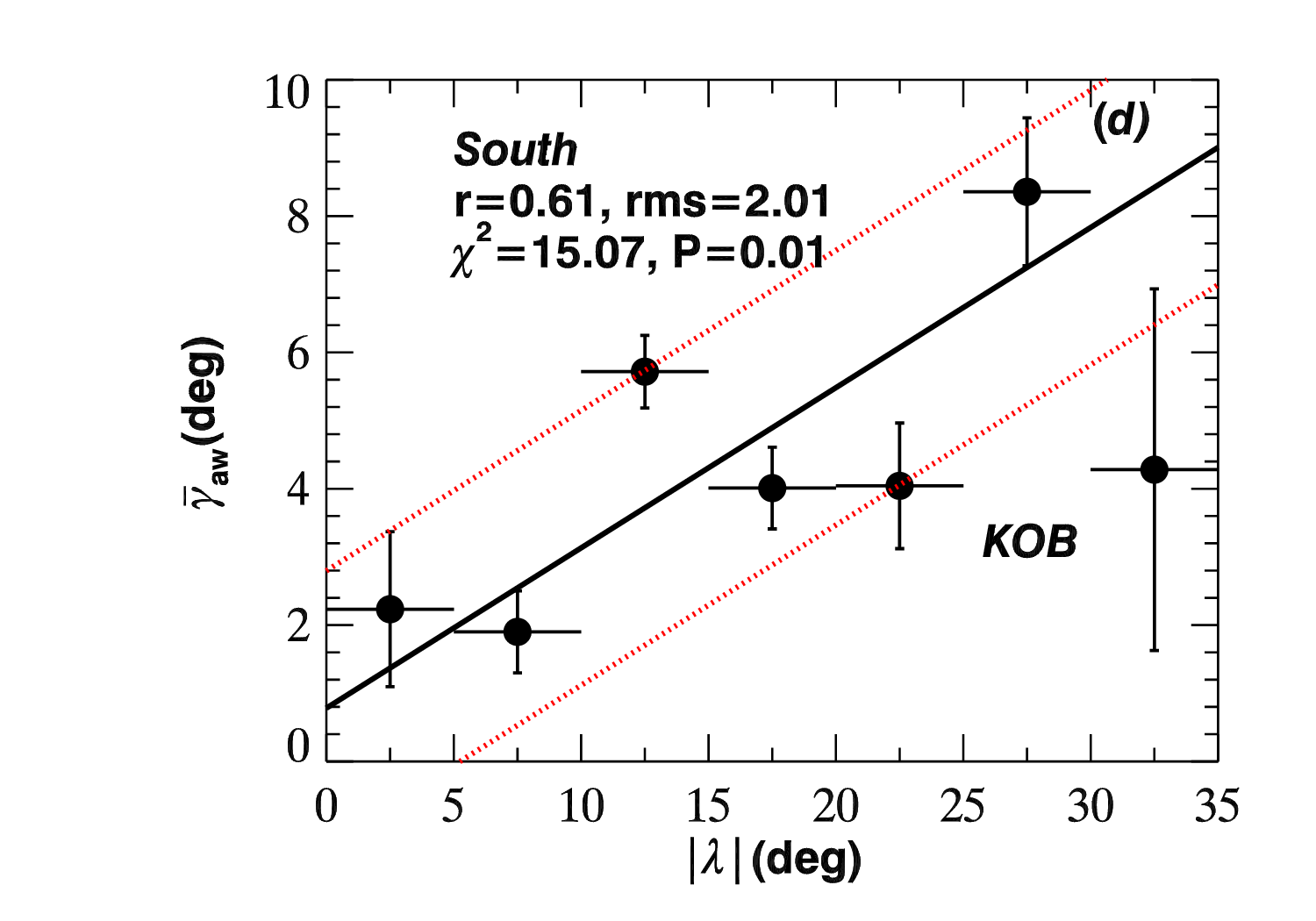}
\caption{Mean tilt angle [$\bar \gamma$] and mean 
 area-weighted tilt angle [$\bar \gamma_{\rm aw}$] versus mid-value
 of the corresponding  $5^\circ$ latitude bin, determined from 
 Mt. Wilson Observatory (MWOB)
 and Kodaikanal Observatory (KOB) southern hemisphere sunspot-group data.  
The {\it continuous line} represents the corresponding  linear-least-square
 best fit and the {\it dotted line} ({\it red}) represents one-rms level.
 The values of the correlation coefficient [$r$], rms, and $\chi^2$ and the
corresponding probability [$P$] are also given.}
\label{f3}
\end{figure}

\begin{table*}
\centering
\caption{Values of the intercept [$c$] and  the slope [$m$], and the 
corresponding standard deviations [$\sigma_c$ and $\sigma_m$], respectively,
  of the best-fit 
linear relationships derived from the mean values of tilt angles 
and area-weighted  tilt angles of sunspot groups  in different
 $5^\circ$-latitude intervals of  
whole-sphere, and northern and southern hemispheres,  by using 
the entire periods MWOB and KOB sunspot-group data.
 The values of correlation coefficient [$r$], 
Student's t [$\tau$] and the corresponding probability [$P$], and
  the ratio $m/\sigma_m$ are also given.}
\begin{tabular}{lccccccccc}
\hline
  \noalign{\smallskip}
$Data$&$c$&$\sigma_c$& $m$ &$\sigma_m$&$r$&$\tau$&$P$&$m/\sigma_m$\\
  \noalign{\smallskip}
\hline
&&&\multicolumn{4}{c}{Derived from  tilt angles}\\
MWOB Whole& 1.26& 0.66 & 0.20& 0.03& 0.97&8.92 &0.0001 & 6.67\\
MWOB North& 1.31& 0.86 & 0.22& 0.05& 0.95&6.80 &0.0005 & 4.40\\
MWOB South& 1.23& 0.83 & 0.17& 0.04& 0.93&5.66 &0.0012 & 4.25\\
KOB Whole & 1.97 & 0.61 & 0.16& 0.03& 0.77&2.70 &0.0214& 5.33\\
KOB North & 2.41 & 0.79 & 0.18& 0.04& 0.93&5.66 &0.0012& 4.50\\
KOB South & 1.47 & 0.83 & 0.15& 0.04& 0.36&0.86 &0.2138& 3.75\\
\\
&&&\multicolumn{4}{c}{Derived from area-weighted tilt angles}\\
MWOB Whole& 2.34&0.67& 0.18& 0.04& 0.97& 8.92  & 0.0001&  4.50\\
MWOB North& 2.32&0.89& 0.23& 0.05& 0.94& 6.16  & 0.0008&  4.60\\
MWOB South& 2.57&0.90& 0.10& 0.05& 0.82& 3.20  & 0.0119&  2.00\\
KOB Whole & 2.39&0.81& 0.20& 0.04& 0.79& 2.88  & 0.0173&  5.00\\
KOB North & 4.15&1.03& 0.13& 0.06& 0.84& 3.46  & 0.0090&  2.17\\
KOB South & 0.78&0.99& 0.24& 0.05& 0.61& 1.72  & 0.0729&  4.80\\
\hline
  \noalign{\smallskip}
\end{tabular}
\label{table2}
\end{table*}

\subsection{Recovering Joy's law From the Data of Individual Solar Cycles}
In this section we find no significant 
results from KOB data.  Here we present 
only the  results derived from MWOB data. We do not know 
the exact reason for the discrepancies in the results derived from KOB 
and MOB data, but  during some 
solar cycles the  KOB data seem to be
 more inconsistent due to a large missing observations~\citep{rcj21}.

In Table~3 we have given the  
values of the intercept [$c$],   slope [$m$], and the
corresponding standard deviation [$\sigma$]  of the best-fit
linear relationships derived from the mean values  
of tilt angles and area-weighted 
 tilt angles of sunspot groups in different  $5^\circ$ absolute latitude intervals of 
the whole sphere  during each
solar cycle. The values of the correlation coefficient $r$,
$\chi^2$, and the corresponding probability [$P$], and  the ratio $m/\sigma_m$
are also given in this table. The corresponding results determined from 
northern and southern hemispheres' data are given in Tables~4 and 5. 
The  best-fit linear relations obtained from the data of some cycles 
are found to be reasonably good (the value of $m$
is statistically significant, i.e. the  ratio $m/\sigma_m$ is 
 $\ge 2$) and the values of $\chi^2$ are reasonably 
small ($P$ is much larger than 0.05) 
and the fits of some other cycles are found to be not good.  

  Figure~4 shows cycle-to-cycle variations in different parameters:
whole sphere's slope [$m_{\rm W}$], northern hemisphere's
slope [$m_{\rm N}$], southern hemisphere's slope
 [$m_{\rm S}$], and  north--south asymmetry in slope  
 [$m_{\rm N}- m_{\rm S}$], determined from the
 MWOB tilt-angle data and the area-weighted tilt-angle data. 
For the sake of comparison in this figure, the variation  in the amplitude 
[$R_{\rm M}$] of the solar cycle 
is also shown. Note that  $m_{\rm W}$,  $m_{\rm N}$, and  $m_{\rm S}$ 
represent the values of the slope $m$ 
of whole sphere, northern hemisphere, and southern hemisphere that 
are given in Tables~3, 4, and 5, respectively.  
 The uncertainty in the  north--south asymmetry $m_{\rm N}- m_{\rm S}$
 is the square-root of 
 the ratio of the sum of the squares of the standard
 deviations of $m_{\rm N}$ and $m_{\rm S}$  to  the number of data points
 (number of absolute latitude bins: 7).  As we can see in this figure,  
the patterns of $m_{\rm W}$ and  $m_{\rm N}$ are similar and both seem to have 
no significant correlation with the amplitude of the solar cycle. 
It looks to be there exists an anti-correlation between $m_{\rm N}$
 and $m_{\rm S}$ (mainly in Figure~4a).
There is a  suggestion of a strong anti-correlation between $m_{\rm S}$ and 
the amplitude of solar cycle, and
  a strong correlation between the north--south 
asymmetry in the slope and the amplitude of the solar cycle 
(mainly in Figure~4b).

 Figure~5 shows the relationship between slope $m_{\rm W}$ (the values 
 of $m$ given Table~3)
  of Joy's law  derived from the whole-sphere
area-weighted tilt-angle data of a solar cycle and 
the  amplitude [$R_{\rm M}$] of the solar cycle. The correlations are  
determined with and  without  the data point of Solar Cycle~15
because it has a high value and is an outlier (more outside
the one-rms level, see Figure~5a).    
As can be seen in this figure there exists  
a small and insignificant anti-correlation and 
 the corresponding best-fit linear relations are not good. That is,  
the slopes of the  best-fit  linear equations  
 are found to be only 1\,--\,1.4 times the corresponding 
standard deviations, although 
 the values of  $\chi^2$  are smaller than that of  the 5\%  significant
level. In the case of tilt-angle data (without area weighting) the 
correlations found to be much smaller (hence not shown).
Overall, these results suggest that there 
exists only a weak  linear relationship between $m$ and $R_{\rm M}$ of
a solar cycle  in the case of the whole sphere data. 
 However, \cite{jiao21} have found the existence of a significant 
anti-correlation between the coefficient of Joy's law and the strength of
the solar cycle by using the area-weighted tilt-angle data of the
 sunspot groups 
with polarity angular separation $> 2.5^\circ$.

  Figure~6 shows the
 relationship  between $m_{\rm N}$ 
(the values of $m$ given in Table~4) 
 obtained from tilt-angle data, and also  that obtained  from area-weighted 
tilt-angle data,  in the  northern hemisphere  during  a solar cycle 
and $R_{\rm M}$ of the solar cycle.
As can be seen in this figure, there exist no significant correlations and 
the corresponding best-fit linear relations are not good.
 That is,
 the value of $r$ is statistically insignificant. The rms is
 large, but it could be mainly due to the data point of Solar Cycle~15, which  
is an outlier. In the case of tilt-angle data, the  $\chi^2$  is significant on
 more than 95\% confidence level 
(probability $P$ is smaller than 0.05). In the case of the area-weighted 
tilt-angle data, the value of $\chi^2$ is reasonably small but the
 ratio $m/\sigma_m$ is found to be significantly small. 
Overall, there 
exists no significant linear relationship between $m_{\rm N}$ and
 $R_{\rm M}$ of 
a solar cycle determined from  the northern hemisphere's data.  
(The values of the slopes are positive, whereas the values of $r$ are negative, 
 see  Figure~6. This discrepancy is due to in  the linear-least-square fit
 calculations uncertainties in abscissae and ordinates were taken into account, 
whereas in the calculations of $r$ the uncertainties are not taken into
 account).

 Figure~7 is the same as  Figure~6, but determined from the southern
 hemisphere's
 data (the values of $m$ given in Table~5). As we can  see in this 
 figure $m_{\rm S}$ is correlated 
 with $R_{\rm M}$ reasonably well 
in  both the cases of the tilt-angle data and  the area-weighted tilt-angle 
data, and  we obtained the following linear relations.

In the case of tilt-angle data:
$$m_{\rm S} = 0.61\pm0.17 - (0.002\pm0.00085) R_{\rm M}, \ {\rm and}  \eqno(2)$$ 
in the case of area weighted tilt-angle data: 
$$m_{\rm S} = 0.84\pm0.21 - (0.0035\pm0.001) R_{\rm M}.  \eqno(3)$$ 

These best-fit linear Equations~2 and 3  are reasonably good, i.e. 
the slopes  are about 2.3 and 3.5 times larger than the corresponding 
standard deviations, respectively. The $P$-values of the corresponding 
correlations are 0.004 and 0.01, i.e. the correlations are significant 
on more than 95\%  confidence level. The corresponding  $\chi^2$-values are 
reasonably 
small, i.e. much lower than the value (12.592) of the 5\% level of significance.
 Moreover,
 only one data point is only slightly outside one-rms level.  
Overall, there 
exists a significant linear relationship between the slope [$m_{\rm S}$]
 determined from 
the southern hemisphere's data  and $R_{\rm M}$.  

Figure 8 shows the
 relationship between  north--south asymmetry in the slope
[$m_{\rm N} - m_{\rm S}$] of a solar cycle and the amplitude
[$R_{\rm M}$] of the solar cycle, for both the cases of
tilt angles and  area-weighted tilt angles (see also Figure~4).
 We obtained the following linear relations.

In the case of tilt angles:
$$m_{\rm N} - m_{\rm S} = -0.60\pm0.11 + (0.0031 \pm 0.0005) R_{\rm M}, \ {\rm and}\eqno(4)$$ 
in the case of area-weighted tilt angles: 
$$m_{\rm N} - m_{\rm S} = -0.63\pm0.11 + (0.0036 \pm 0.0005) R_{\rm M}.  \eqno(5)$$ 

In the case of Equation~4, the corresponding correlation is insignificant and 
$\chi^2$ is very large mainly because the data point of Solar Cycle~15 is 
 far away from the one-rms level. However, the slope is  6.2 times larger than 
the corresponding standard deviation, suggesting a possibility of the 
existence of the linear relation between   $m_{\rm N} - m_{\rm S}$ 
and $R_{\rm M}$ of a solar cycle.
 In the case of Equation~5 the correlation is highly 
significant (Student's t is 5.9 and the corresponding $P = 0.001$), the 
$\chi^2$ is considerably small (the corresponding $P = 0.61$), only 
one data point is only slightly out of the one-rms, and  moreover the 
slope is  7.2 times larger than the corresponding standard deviation.  
Overall, we find that there  exists a good linear relationship  between the
 north--south difference in the slope and the  amplitude [$R_{\rm M}$] of
 a solar cycle.

\begin{table*}
\centering
\caption{Whole sphere:
 Values of the intercept [$c$] and the  slope [$m$], and the 
corresponding standard deviations  [$\sigma_c$ and $\sigma_m$],
 respectively,  of the best-fit linear relationships derived from the mean
 values of tilt angles and area-weighted  tilt angles of sunspot groups in
 different $5^\circ$-latitude intervals of the whole-sphere
 during each solar cycle (SC). The values of correlation coefficient [$r$], 
$\chi^2$ and the corresponding probability [$P$], the ratio $m/\sigma_m$,
 and the rms  are also given. $^{\mathrm a}$ indicates incomplete data.}
\begin{tabular}{lccccccccc}
\hline
  \noalign{\smallskip}
SC&$c$&$\sigma_c$& $m$ &$\sigma_m$&$r$&$\chi^2$&
$P$&$m/\sigma_m$&rms\\
  \noalign{\smallskip}
\hline
&&&\multicolumn{4}{c}{Derived from  tilt angles}\\
 15$^{\mathrm a}$& $-3.29$&  2.41&  0.55&  0.15&  0.82&  5.38&  0.25& 3.67&  2.93\\
 16&  0.69&  1.84&  0.25&  0.10&  0.87&  5.89&  0.32&  2.50&  1.63\\
 17&  2.29&  1.48&  0.18&  0.08&  0.53&  4.58&  0.47&  2.25&  1.62\\
 18&  1.79&  1.32&  0.18&  0.07&  0.21&  8.94&  0.11&  2.57&  2.90\\
 19&  0.21&  1.20&  0.20&  0.06&  0.95&  3.19&  0.67&  3.33&  0.75\\
 20&  1.50&  1.38&  0.18&  0.07&  0.81&  4.00&  0.55&  2.57&  1.21\\
 21&  1.47&  1.45&  0.22&  0.08&  0.95&  1.45&  0.92&  2.75&  0.72\\
\\
&&&\multicolumn{4}{c}{Derived from area-weighted tilt angles}\\
 15$^{\mathrm a}$ & -1.36&  2.57&  0.53&  0.16&  0.81&  7.50&  0.11& 3.31&  2.79\\
 16&  2.65&  1.74&  0.17&  0.09&  0.82&  4.05&  0.54&  1.89&  2.02\\
 17&  2.39&  1.55&  0.24&  0.08&  0.55&  9.21&  0.10&  3.00&  2.60\\
 18&  3.13&  1.49&  0.16&  0.08&  0.01& 10.68&  0.06&  2.00&  3.51\\
 19&  1.35&  1.18&  0.15&  0.06&  0.84&  6.46&  0.26&  2.50&  0.98\\
 20&  0.18&  1.55&  0.32&  0.08&  0.92&  2.69&  0.75&  4.00&  1.24\\
 21&  3.00&  1.58&  0.16&  0.08&  0.54&  6.86&  0.23&  2.00&  2.21\\
\hline
  \noalign{\smallskip}
\end{tabular}
\label{table3}
\end{table*}

\begin{table*}
\centering
\caption{Northern hemisphere:
 Values of the intercept [$c$] and  the slope [$m$],  and the 
corresponding standard deviations [$\sigma_c$ and $\sigma_m$], respectively,  
 of the best-fit 
linear relationships derived from the mean values of tilt angles 
 and  area-weighted 
 tilt angles of sunspot groups in different $5^\circ$-latitude intervals of the northern  
hemisphere during each 
solar cycle (SC). The values of correlation coefficient [$r$], 
$\chi^2$ and the corresponding probability [$P$], the ratio $m/\sigma_m$, 
and the rms  are also given. $^{\mathrm a}$ indicates incomplete data.}
\begin{tabular}{lccccccccc}
\hline
  \noalign{\smallskip}
SC&$c$&$\sigma_c$& $m$ &$\sigma_m$&$r$&$\chi^2$&
$P$&$m/\sigma_m$&rms\\
  \noalign{\smallskip}
\hline
&&&\multicolumn{4}{c}{Derived from  tilt angles}\\
 15$^{\mathrm a}$&$ -5.70$& 3.45&  0.79&  0.21&  0.91& 4.64&  0.33&  3.76&  2.98\\
 16&  1.43&  2.24&  0.18&  0.12&  0.78&  3.27&  0.66&  1.50&  2.12\\
 17&  5.00&  2.05&  0.05&  0.11&  0.03&  5.60&  0.35&  0.45&  2.05\\
 18&  3.93&  1.86&  0.14&  0.10&  0.31&  6.15&  0.29&  1.40&  3.50\\
 19&  0.19&  1.47&  0.27&  0.08&  0.96&  2.09&  0.84&  3.38&  0.88\\
 20&  1.24&  1.82&  0.15&  0.10&$-0.13$& 2.92&  0.71&  1.50&  2.94\\
 21&  0.13&  2.04&  0.34&  0.11&  0.69&  6.88&  0.23&  3.09&  2.76\\
\\
&&&\multicolumn{4}{c}{Derived from area weighted-tilt angles}\\
 15$^{\mathrm a}$&$ -0.86$& 3.09& 0.49&  0.19&  0.82&  6.52&  0.16&  2.58&  2.72\\
 16&  2.20&  2.14&  0.15&  0.11&  0.48&  6.08&  0.30&  1.36&  3.23\\
 17&  3.68&  2.31&  0.25&  0.12&  0.14& 10.20&  0.07&  2.08&  6.92\\
 18&  4.84&  1.91&  0.13&  0.10&  0.06&  9.34&  0.10&  1.30&  4.30\\
 19&  1.24&  1.37&  0.25&  0.07&  0.94&  1.87&  0.87&  3.57&  0.88\\
 20&  0.96&  1.95&  0.25&  0.10&  0.14&  2.64&  0.75&  2.50&  4.23\\
 21&  2.57&  2.19&  0.25&  0.12&  0.20& 12.60&  0.03&  2.08&  3.68\\
\hline
  \noalign{\smallskip}
\end{tabular}
\label{table4}
\end{table*}

\begin{table*}
\centering
\caption{Southern hemisphere: Values of the intercept [$c$] and the slope [$m$],
 and the corresponding standard deviations  [$\sigma_c$ and  $\sigma_m$], 
respectively,  of the best-fit 
linear relationships derived from the mean values of tilt angles 
 and  area-weighted 
 tilt angles of sunspot groups   in different $5^\circ$-latitude intervals of
 the southern 
hemisphere during each 
solar cycle (SC). The values of the correlation coefficient [$r$], 
$\chi^2$ and the corresponding probability [$P$], the ratio $m/\sigma_m$, 
and the rms are also given. $^{\mathrm a}$ indicates incomplete data.} 

\begin{tabular}{lccccccccc}
\hline
  \noalign{\smallskip}
SC&$c$&$\sigma_c$& $m$ &$\sigma_m$&$r$&$\chi^2$&
$P$&$m/\sigma_m$&rms\\
  \noalign{\smallskip}
\hline
&&&\multicolumn{4}{c}{Derived from  tilt angles}\\
 15$^{\mathrm a}$& $-0.33$&  3.06&  0.31&  0.19&  0.38&  3.67&  0.45& 1.63&  2.88\\
 16&  0.09&  2.70&  0.31&  0.14&  0.43&  8.00&  0.16&  2.21&  4.57\\
 17& $-0.18$&  2.00&  0.30&  0.10&  0.57&  4.77&  0.44&  3.00&  2.61\\
 18&  0.87&  1.77&  0.15&  0.09& $-0.03$& 10.60&  0.06&  1.67&  3.08\\
 19&  1.62&  1.51&  0.02&  0.08&  0.54&  1.80&  0.88&  0.25&  1.10\\
 20&  1.57&  2.04&  0.24&  0.11&  0.64&  6.40&  0.27&  2.18&  2.57\\
 21&  2.25&  2.04&  0.15&  0.11&  0.65&  5.72&  0.33&  1.36&  1.97\\
\\
&&&\multicolumn{4}{c}{Derived from area weighted-tilt angles}\\
 15$^{\mathrm a}$&  0.73&  3.81&  0.38&  0.24&  0.69&  4.54&  0.34& 1.58&  2.76\\
 16&  2.36&  2.67&  0.25&  0.14&  0.34&  3.84&  0.57&  1.79&  3.59\\
 17&  1.11&  1.79&  0.21&  0.09&  0.32&  4.04&  0.54& 2.33&  2.69\\
 18&  4.18&  1.87&$ -0.03$&  0.10& $-0.28$& 11.47& 0.04&$ -0.30$&  2.82\\
 19&  3.82&  1.94&$ -0.14$&  0.10& $-0.56$&  5.33& 0.38&$ -1.40$&  1.39\\
 20& $-0.92$&  2.21&  0.40&  0.11&  0.87&  4.90&  0.43& 3.64&  2.02\\
 21&  3.64&  2.57&  0.09&  0.14&  0.24& 15.14&  0.01& 0.64&  5.06\\
\hline
  \noalign{\smallskip}
\end{tabular}
\label{table5}
\end{table*}

\begin{figure}
\centering
\includegraphics[width=6.0cm]{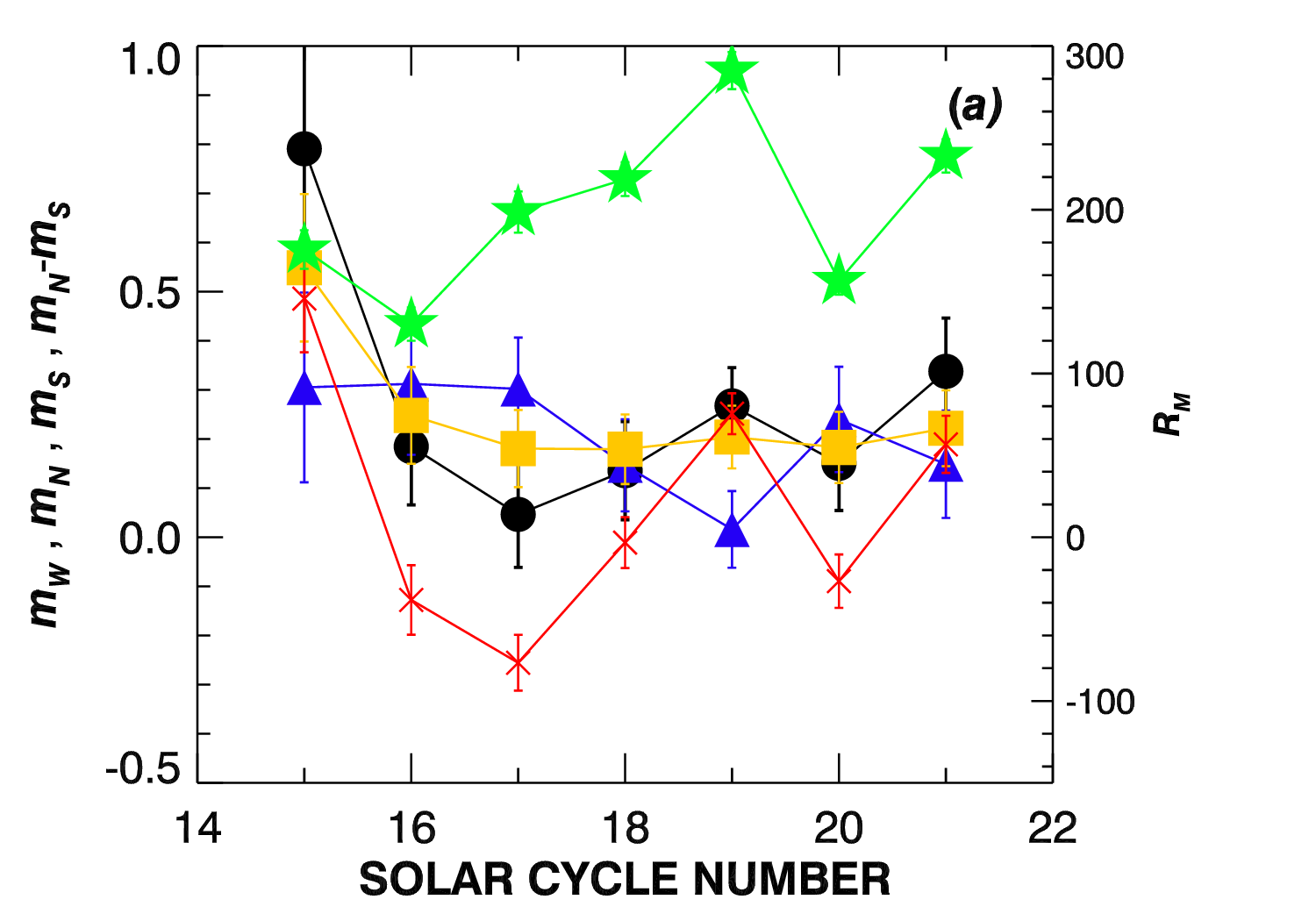}
\includegraphics[width=6.0cm]{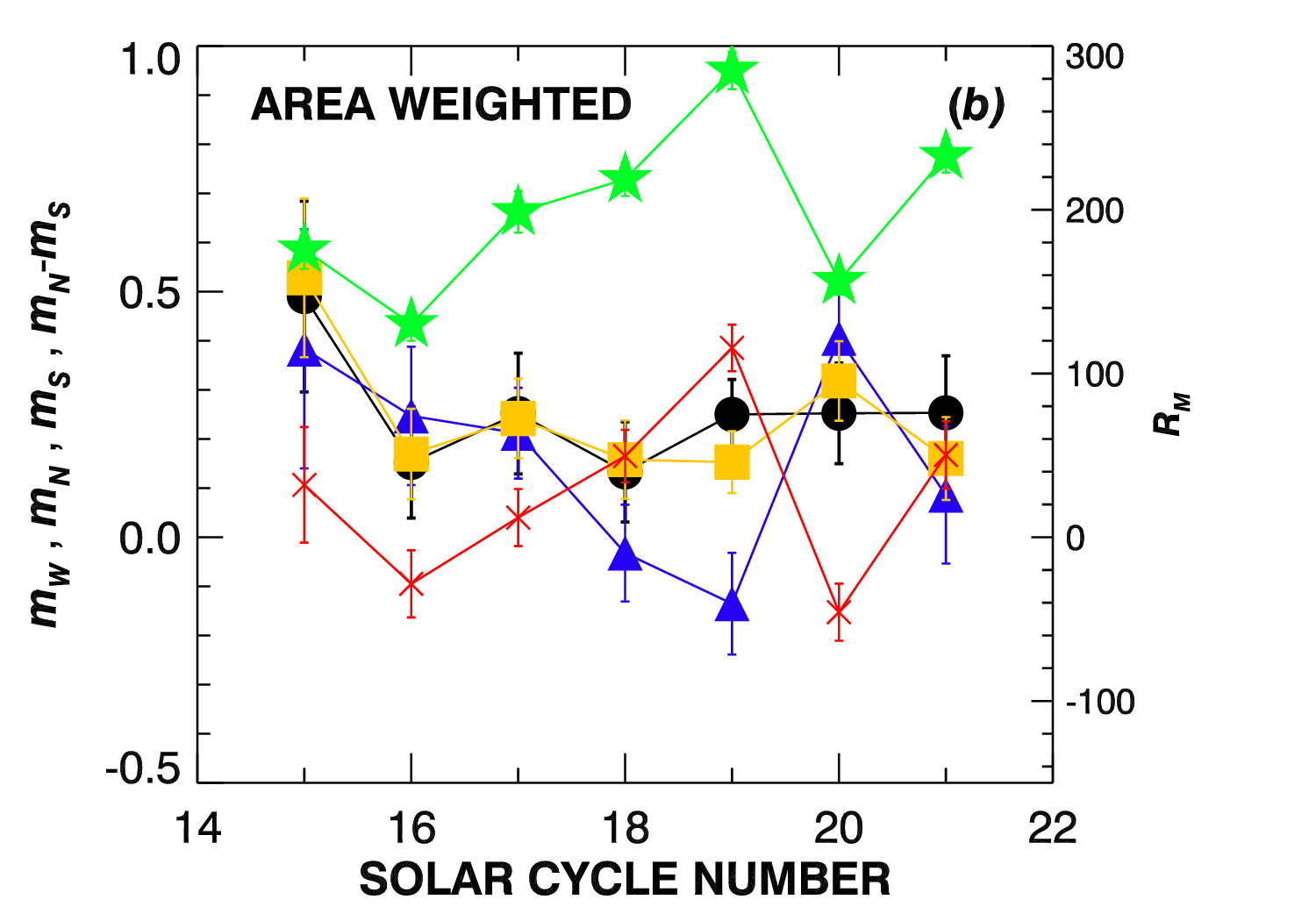}
\caption{Cycle-to-cycle variations in different parameters:
whole sphere's slope $m_{\rm W}$ ({\it yellow filled square}), 
northern hemisphere's
slope $m_{\rm N}$ ({\it black filled circle}), southern hemisphere's slope  
$m_{\rm S}$ ({\it blue filled triangle}), and  north--south asymmetry in  
slope  $m_{\rm N}- m_{\rm S}$ ({\it red cross}).
 ({\bf a}) Determined from MWOB tilt-angle data and 
       ({\bf b}) determined from MWOB area-weighted tilt-angle data. Note 
that the  data of Solar Cycle~15 are incomplete. 
The variation in  $R_{\rm M}$ ({\it green filled star}) is also shown.}
\label{f4}
\end{figure}

\begin{figure}
\centering
\includegraphics[width=6.0cm]{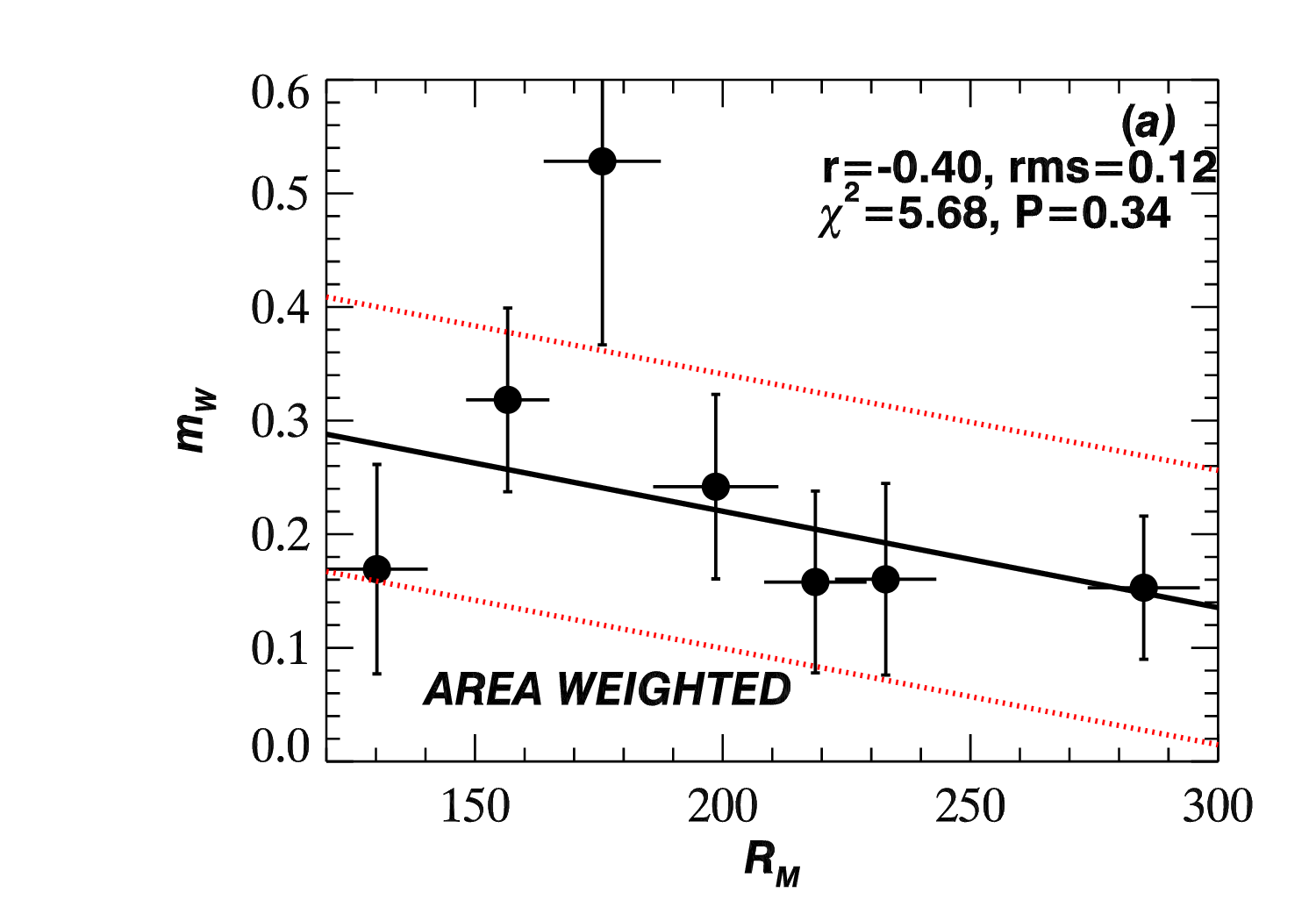}
\includegraphics[width=6.0cm]{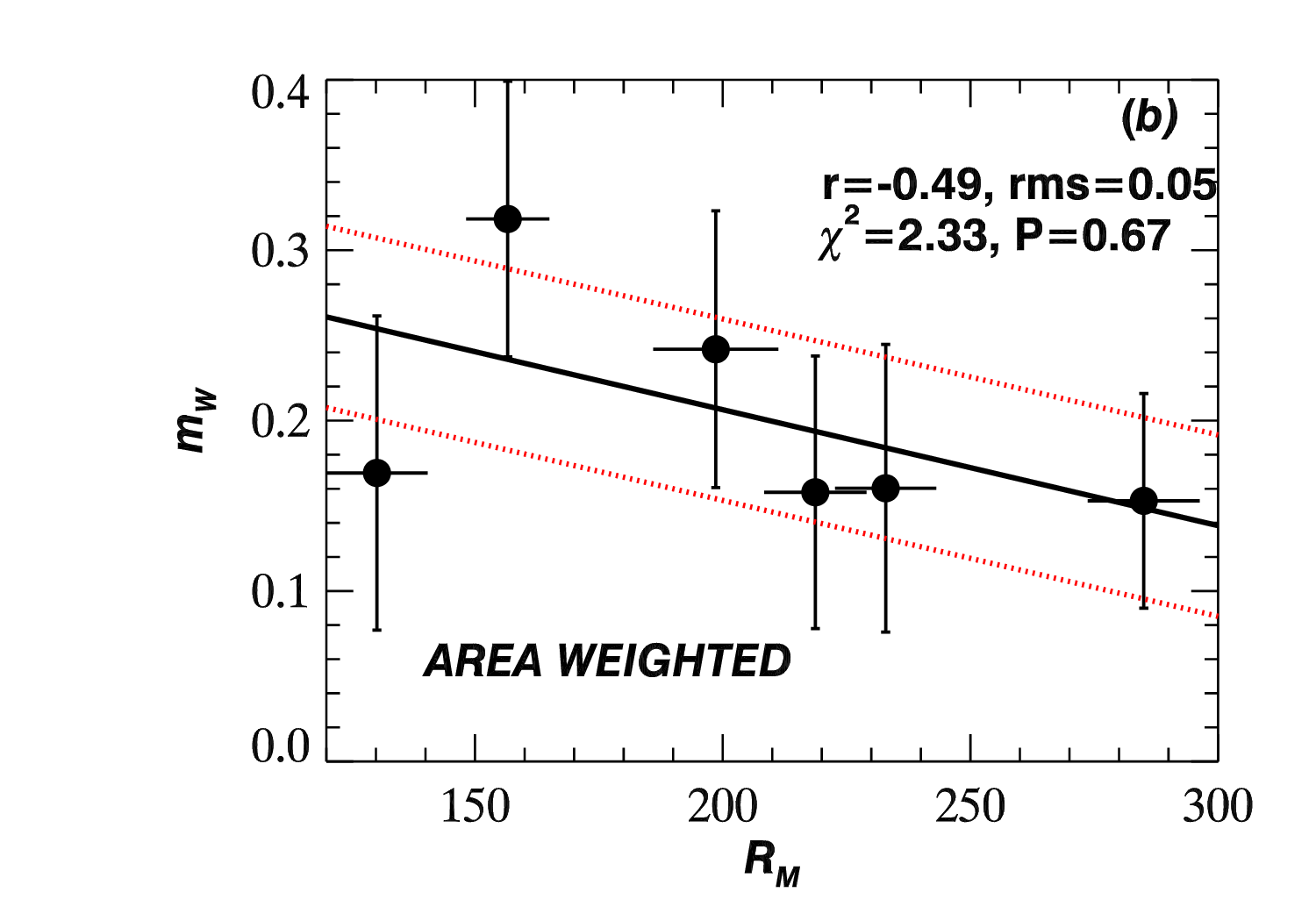}
\caption{The slope [$m_{\rm W}$]  of Joy's law derived from the whole sphere 
area-weighted tilt-angle data  in $5^\circ$ absolute latitude bins
 versus amplitude  [$R_{\rm M}$]  of solar cycle, 
 ({\bf a}) for all Cycles 15\,--\,21 and 
       ({\bf b}) for Cycles 16\,--\,21.
The {\it continuous line} represents the corresponding 
 linear-least-square best fit
 and the {\it dotted line} ({\it red}) represents one-rms level.
 The values of the
 correlation coefficient [$r$], rms, and $\chi^2$ and the
corresponding probability [$P$] are also given.}
\label{f5}
\end{figure}

\begin{figure}
\centering
\includegraphics[width=6.0cm]{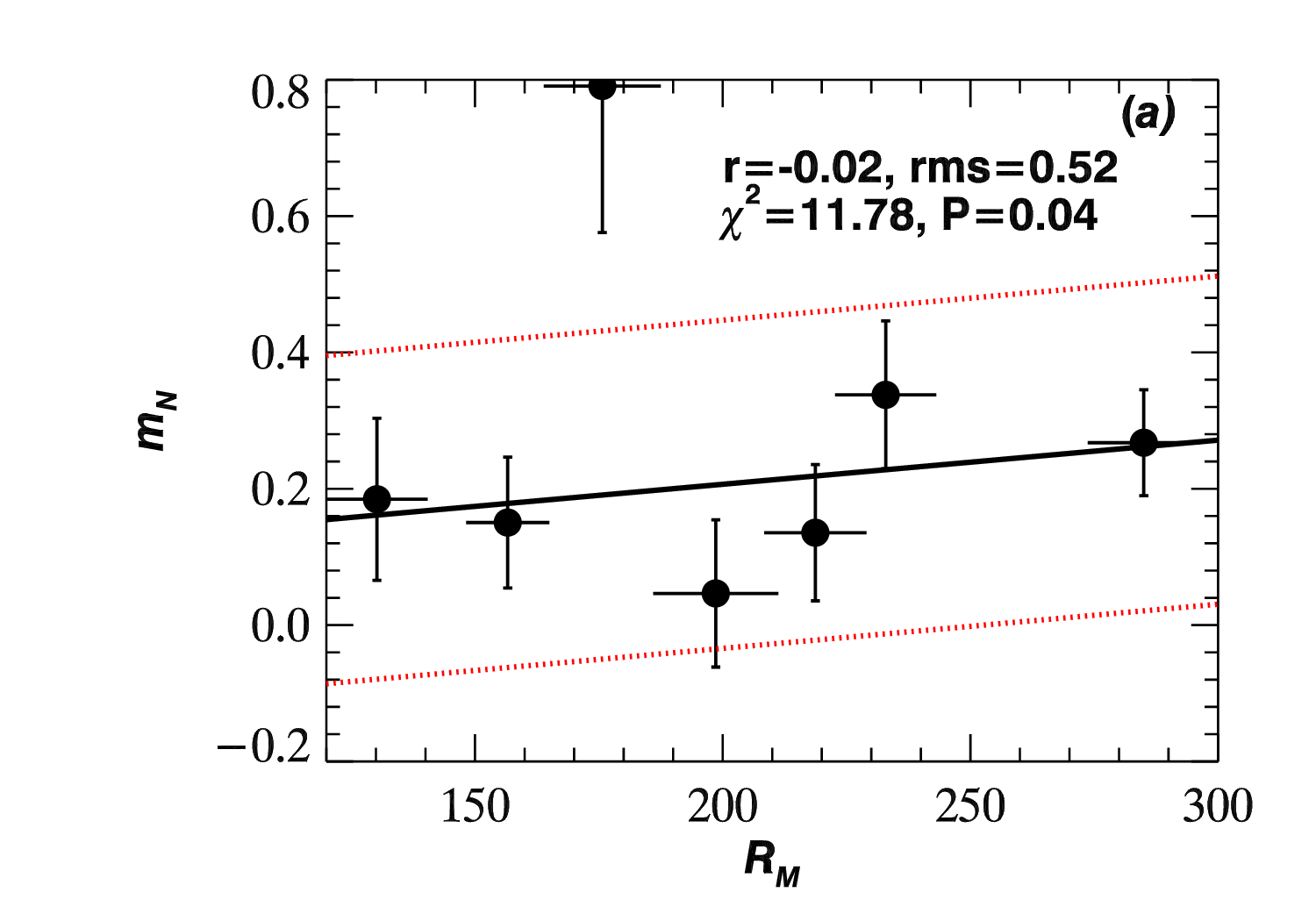}
\includegraphics[width=6.0cm]{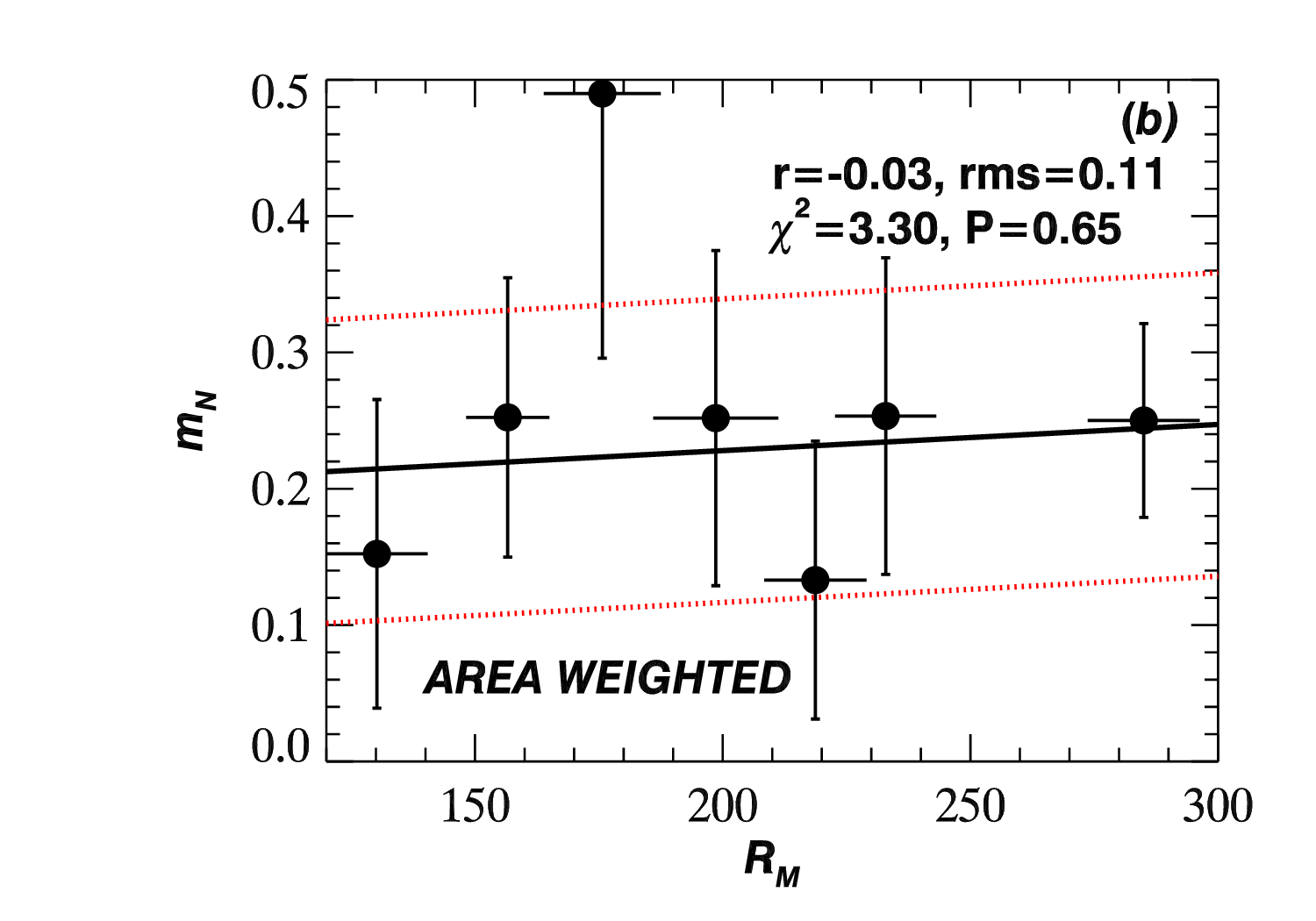}
\caption{The slope [$m_{\rm N}$]  of Joy's law in northern hemisphere 
 versus amplitude  [$R_{\rm M}$] of solar cycle  derived from
 the average values of  ({\bf a}) tilt angles, and 
       ({\bf b}) area-weighted tilt angles, in $5^\circ$ latitude bins.
The {\it continuous line} represents the corresponding  linear-least-square
 best fit and the {\it dotted line} ({\it red}) represents one-rms level.
 The values of the  correlation coefficient [$r$], rms, and $\chi^2$ and the 
corresponding probability [$P$] are also given.}   
\label{f6}
\end{figure}

\begin{figure}
\centering
\includegraphics[width=6.0cm]{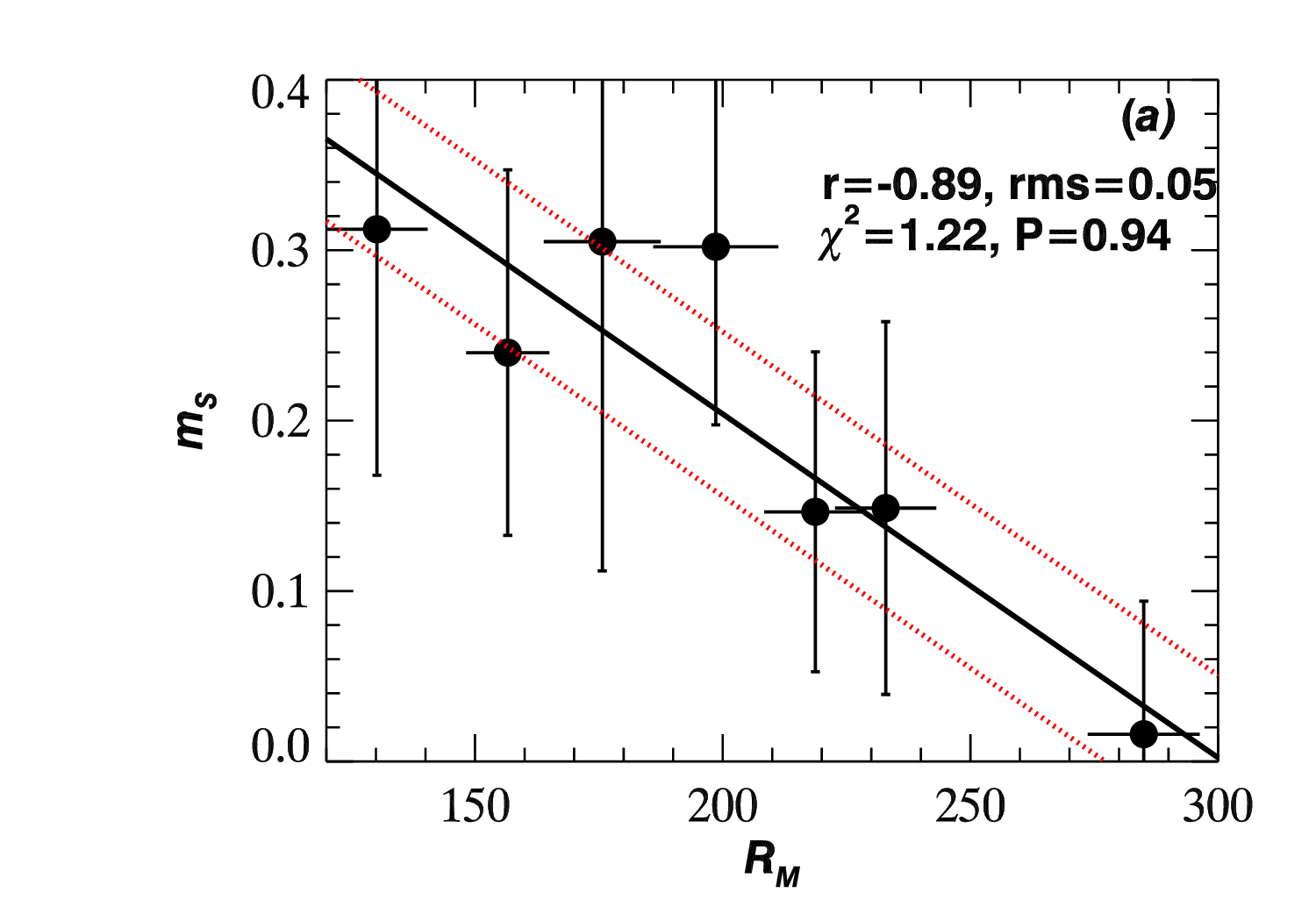}
\includegraphics[width=6.0cm]{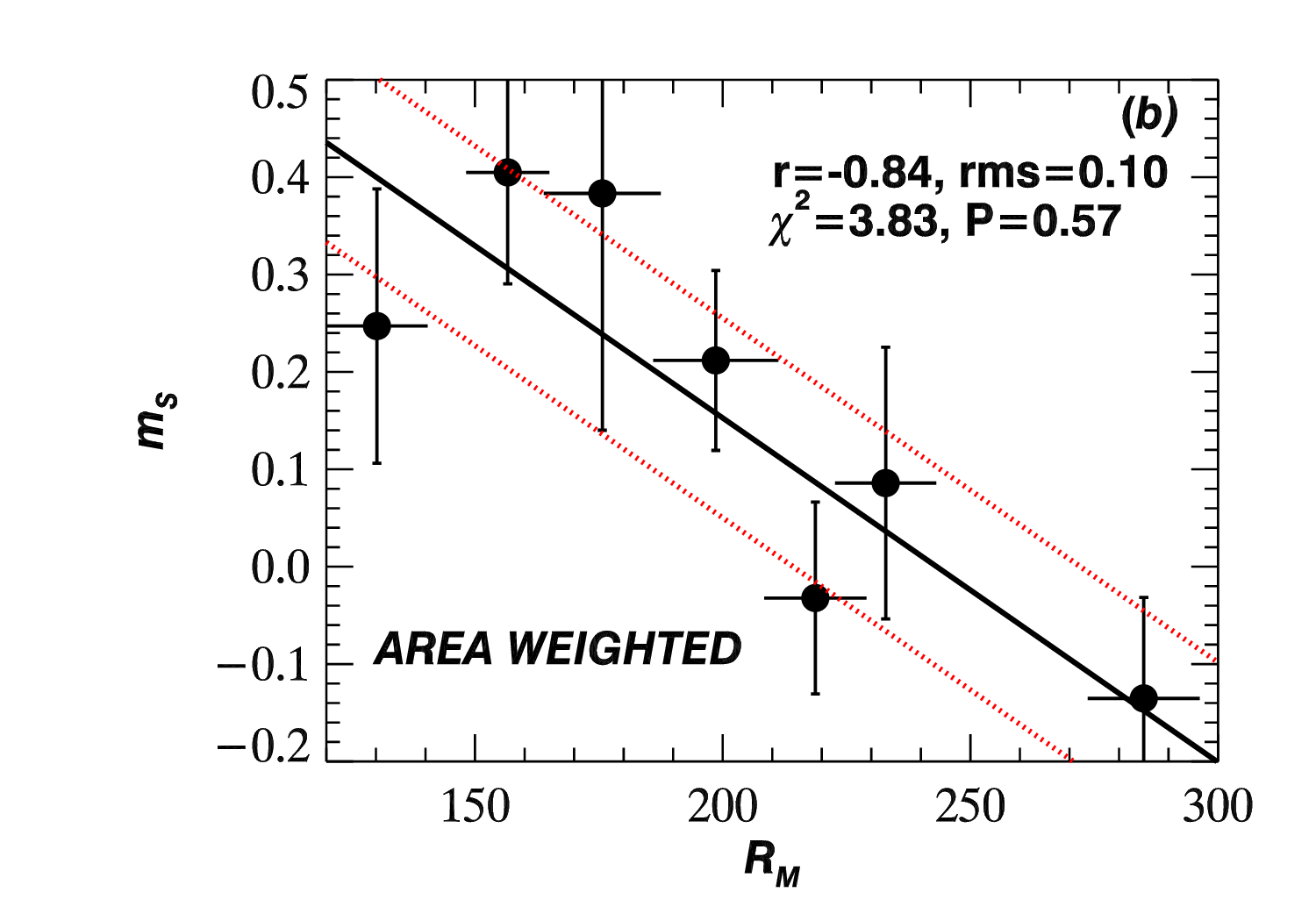}
\caption{The slope [$m_{\rm S}$]  of Joy's law in southern hemisphere
 versus amplitude  [$R_{\rm M}$] of solar cycle derived from the
 average values of ({\bf a}) tilt angles, and
       ({\bf b})  area-weighted tilt angles, in $5^\circ$ latitude bins.
The {\it continuous line} represents the corresponding  linear-least-square
 best fit  and the {\it dotted line} ({\it red}) represents one-rms level.
 The values of the correlation coefficient [$r$], rms, and $\chi^2$ and the 
corresponding probability [$P$] are also given.}
\label{f7}
\end{figure}

\begin{figure}
\centering
\includegraphics[width=6.0cm]{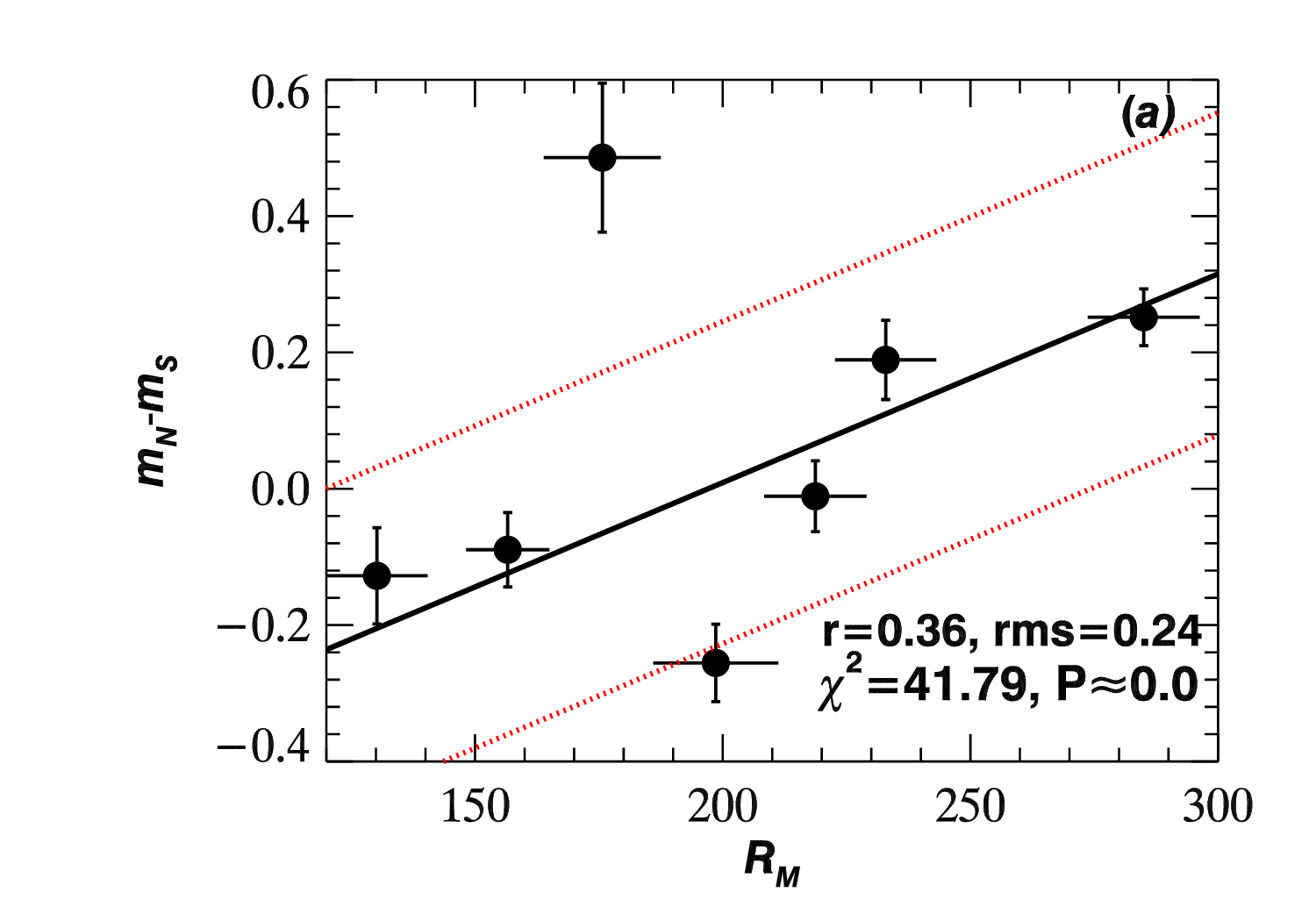}
\includegraphics[width=6.0cm]{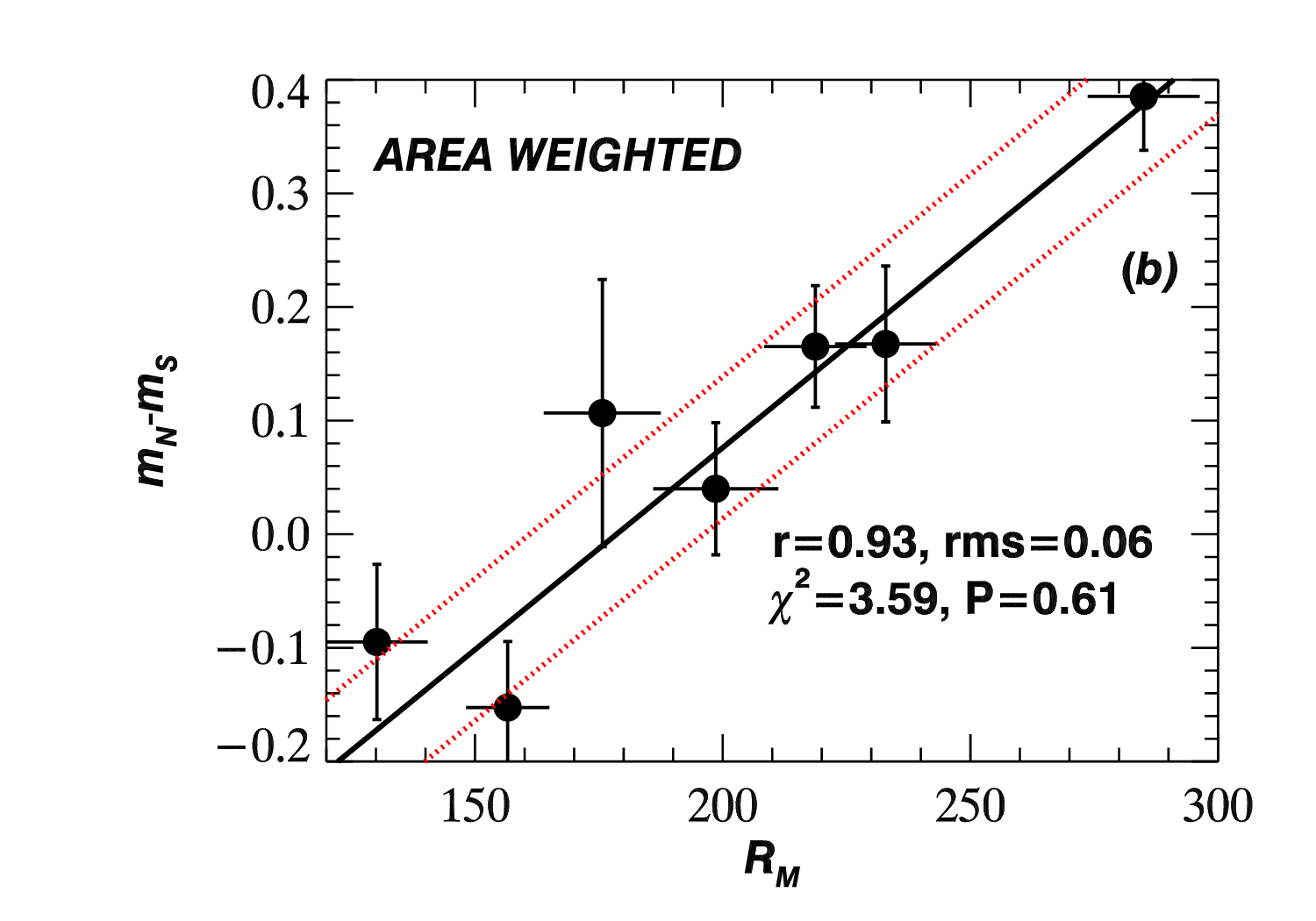}
\caption{The hemispheric difference [$m_{\rm N} - m_{\rm S}$] in the slopes 
  of Joy's law (see also Figure~4),  derived from  ({\bf a})  tilt angles
 and  ({\bf b}) area-weighted tilt angles shown in Figures~6 and 7, 
 versus amplitude  [$R_{\rm M}$] of solar cycle. 
The {\it continuous line} represents the corresponding  linear-least-square
 best fit and the {\it dotted line} ({\it red}) represents one-rms level.
 The values of the correlation coefficient [$r$], rms, and $\chi^2$ and the
corresponding probability [$P$] are also given.}
\label{f8}
\end{figure}

\section{Discussion and Conclusions}
Study of tilt angles of solar bipolar magnetic regions is important because the
tilt angles have an important role in the solar dynamo. We analysed the data
on tilt angles of sunspot groups measured at 
  MWOB during the period 1917\,--\,1986 and
  at KOB during the period 1906--1986.
We have used the amplitudes (the values of $R_{\rm M}$) of 
Solar Cycles 15\,--\,21. 
 We binned the daily tilt-angle data
during each of the Solar Cycles 15\,--\,21  into different
$5^\circ$-latitude intervals and calculated the mean value of the tilt angles
  in each latitude interval and the corresponding standard error. We fitted
these binned data to Joy's law, i.e.
the linear relationship between tilt angle and  latitude of an
active region.
 The linear-least-square fit calculations were done by taking into account
the uncertainties  in both the abscissa (latitude) and
 ordinate (mean tilt angle). 
The calculations were carried out
  by using both the tilt-angle and  area-weighted tilt-angle data on
 the whole sphere, northern hemisphere,  and southern hemisphere
 during the  whole period  and  during each individual solar cycle.
We find a significant
 difference (north--south asymmetry) between the slopes of Joy's law
 recovered from northern and southern hemispheres' whole period MWOB data
of area-weighted tilt angles.
The slope obtained from the whole sphere's MWOB data of a solar cycle
 is found to be weakly anti-correlated  (statistically insignificant)  to the
 amplitude [$R_{\rm M}$] of the solar cycle.  No correlation  is found
 between the slope
obtained from the northern hemisphere's data and the amplitude of solar cycle,
 whereas the slope obtained from the southern hemisphere's  data is found 
to be  reasonably well anti-correlated with the amplitude of solar cycle.
  In the case of area weighted tilt-angle data, highly statistically
 significant correlation is found between the  north--south
 asymmetry in the slope of a solar cycle and the  amplitude of
the  solar cycle.
The corresponding best-fit linear equations are  found to be statistically
 significant. These results are not found in KOB data.

\citet{norton13} 
analysed the Mt.
Wilson sunspot-group data (1917\,--\,1986) and found  that 
 anti-correlation between 
 $\langle \gamma \rangle/\langle |\lambda| \rangle$
  and  strength  of a solar cycle is significant only in southern 
hemisphere.  Here we analysed the same data and 
 find  a significant  anti-correlation between 
the slope and  amplitude $R_{\rm M}$ of solar cycle 
 only from the southern hemisphere's data.
That is, the  behavior of slope  in northern and southern hemispheres is 
  closely similar to that of 
 $\langle \gamma \rangle/\langle |\lambda| \rangle$
  found by \citet{norton13}. 
(Note that the slope $m$ represents the latitudinal gradient of tilt angle, 
whereas 
 $\langle \gamma \rangle/\langle |\lambda| \rangle$
 is considered as the latitude independent mean 
tilt angle.) 
Here we also find the existence of a significant correlation 
between north--south difference in the slope and $R_{\rm M}$ of a solar cycle. 
The reason behind why the correlation between the slope and $R_{\rm M}$ 
 is significant only in southern hemisphere is not clear to us.  
We  find  no significant correlation 
between the slope (or north--south difference $m_{\rm N} - m_{\rm S}$) and  
north--south difference
in the mean area of sunspot groups \citep[taken from][]{jj22} at the epoch of
  $R_{\rm M}$ of a solar cycle, indicating that no correlation exists 
between the former and the north--south asymmetry in the amplitude of the solar
 cycle.

 The $m_{\rm S}$ strongly varies and strongly anti-correlates 
 with  $R_{\rm M}$, but $m_{\rm N}$ is almost constant (if we exclude the 
large value  of Solar Cycle~15, see Figures~4 and 6)
 and there  exists no 
  correlation between $m_{\rm N}$ and $R_{\rm M}$. Hence, there exists only a
 weak anti-correlation between $m_{\rm W}$ (derived from the 
combined northern and southern hemispheres' data) and $R_{\rm M}$
 and a good  positive  correlation exists between 
 $m_{\rm N} - m_{\rm S}$ and $R_{\rm M}$. These results convey 
that it is important to determine the slope of the Joy's law  separately
from the tilt-angle data of the two 
hemispheres. However, the physical
 reason behind why $m_{\rm S}$ strongly
  anti-correlates with $R_{\rm M}$ and no correlation exists between 
$m_{\rm N}$ and $R_{\rm M}$ is not known to us. 
According to \cite{jiao21} the Coriolis force 
involved in the formation of the tilt angle depends on the local
expansion rate of the rising flux tube and the local rotation rate, 
hence the tilt angles of the two hemispheres could be
 independent and uncoupled. As already mentioned in Section~1, there exist 
north--south differences in the  rotational and
  meridional motions of sunspot groups and also there exist 
cycle-to-cycle variations in these differences \citep[e.g.][]{ju06}. 
The differences in the cycle-to-cycle variations in the  rotational 
and meridional motions of sunspot groups in northern 
and southern hemispheres may be responsible for the hemispheric differences in 
the cycle-to-cycle  variations in 
the coefficient of Joy's law.  
 It needs to be investigated.

\begin{acks}
 The author thanks the anonymous reviewer for useful comments and 
suggestions.
The author acknowledges the work of all the
 people who contribute to and maintain the MWOB and KOB  sunspot databases.
 We have used the maximum values of Solar Cycles 15\,--\,21
 determined by \citet{pesnell18}
from the time series of  13-month smoothed monthly mean
values of  version~2 of international sunspot number (SN) available
 at {\sf www.sidc.be/silso/datafiles}.
\end{acks}

\begin{ethics}
 \begin{conflict}
The author declares that he has no conflicts of interest.
 \end{conflict}
 \end{ethics}

\begin{dataavailability}
 All data generated or analysed during this study are included in this article.
\end{dataavailability}
 \bibliographystyle{spr-mp-sola}
 \bibliography{tiltbib.bib}

\end{article}
\end{document}